\UseRawInputEncoding
\documentclass[superscriptaddress,twocolumn,amsmath,amssymb,aps,prx]{revtex4-2}

\usepackage{graphicx}
\usepackage{soul}
\usepackage{siunitx}
\usepackage{makecell}
\usepackage{hyperref}
\usepackage[capitalize]{cleveref}
\begin{document}

\title{Active Fourier Spectroscopy:\\From fixed sampling protocols to adaptive measurement}

\author{J.~Schr\"oder}
\email{jakob.schroeder@physik.uni-muenchen.de}
\affiliation{Fakult\"at f\"ur Physik, Ludwig-Maximilians-Universit\"at M\"unchen, Am Coulombwall 1, 85748 Garching, Germany}
\affiliation{Munich Center for Machine Learning (MCML)}

\author{S.~Howard}
\affiliation{Fakult\"at f\"ur Physik, Ludwig-Maximilians-Universit\"at M\"unchen, Am Coulombwall 1, 85748 Garching, Germany}
\affiliation{Munich Center for Machine Learning (MCML)}
\affiliation{Department of Physics, Clarendon Laboratory, University of Oxford, Oxford OX1 3PU, United Kingdom}

\author{C.~Eberle}
\affiliation{Fakult\"at f\"ur Physik, Ludwig-Maximilians-Universit\"at M\"unchen, Am Coulombwall 1, 85748 Garching, Germany}
\affiliation{Munich Center for Machine Learning (MCML)}

\author{J.~Esslinger}
\affiliation{Fakult\"at f\"ur Physik, Ludwig-Maximilians-Universit\"at M\"unchen, Am Coulombwall 1, 85748 Garching, Germany}

\author{N.~Leopold-Kerschbaumer}
\affiliation{Fakult\"at f\"ur Physik, Ludwig-Maximilians-Universit\"at M\"unchen, Am Coulombwall 1, 85748 Garching, Germany}
\affiliation{Max Planck Institut f\"ur Quantenoptik, Hans-Kopfermann-Strasse 1, Garching 85748, Germany}
\affiliation{Center for Molecular Fingerprinting (CMF), Budapest, Hungary}

\author{K.~V.~Kepesidis}
\affiliation{Fakult\"at f\"ur Physik, Ludwig-Maximilians-Universit\"at M\"unchen, Am Coulombwall 1, 85748 Garching, Germany}
\affiliation{Max Planck Institut f\"ur Quantenoptik, Hans-Kopfermann-Strasse 1, Garching 85748, Germany}
\affiliation{Center for Molecular Fingerprinting (CMF), Budapest, Hungary}

\author{A.~D\"opp}
\affiliation{Fakult\"at f\"ur Physik, Ludwig-Maximilians-Universit\"at M\"unchen, Am Coulombwall 1, 85748 Garching, Germany}
\affiliation{Munich Center for Machine Learning (MCML)}
\affiliation{Max Planck Institut f\"ur Quantenoptik, Hans-Kopfermann-Strasse 1, Garching 85748, Germany}

\begin{abstract}
Fourier-transform spectroscopy (FTS) is based on the Nyquist-Shannon theorem, which dictates a strict sampling protocol for the measurement of continuous signals.
Here we introduce \emph{Active Fourier Spectroscopy} (AFS), a generalization of FTS in which measurements are still taken in the Fourier-conjugate delay domain, but where the most informative acquisition points are selected adaptively and the spectrum is reconstructed by sequential, uncertainty-aware inference rather than by fixed-grid Fourier inversion.
Our approach relies on a theoretical insight that the Nyquist condition applies only under complete ignorance about the system under investigation, and shows how to systematically improve upon it when prior knowledge exists. Because AFS recovers conventional FTS whenever the prior is uninformative, its performance never falls below that of conventional sampling.
We demonstrate AFS for optical spectroscopy in three settings vital to research and medicine: clinical FTIR-based blood plasma analysis, optical metrology of structured laser beams, and hyperspectral imaging with an RGB camera. In each case AFS reaches a higher target reconstruction quality from far fewer acquisitions and provides complete uncertainty quantification in real time, propagated rigorously through downstream analysis.
\end{abstract}

\maketitle
\tableofcontents

\section{Introduction}

\noindent

Fourier Transform Spectroscopy (FTS) is one of the most widely-used analytical methods in science, medicine and industry today. Fourier transform infrared (FTIR) spectroscopy \cite{GriffithsFTIR2007}, for instance, is ubiquitous in chemical laboratories \cite{enders_functional_2021, dendisova_use_2018}, industrial quality control \cite{fakayode_molecular_2020}, medical diagnostics \cite{movasaghi2008fourier}, and environmental monitoring \cite{simonescu_application_2012}. At its core, FTS measures the fluence of light, $F$, as a function of the time delay, $\tau$, between two identical copies of the optical signal. Depending on the delay, different wavelengths add constructively or destructively, so that each measurement is a linear projection of the spectrum, and the combined signal forms an autocorrelation trace described by
\begin{equation}
    F(\tau) = \int S(\omega)\bigl(1 + \cos(\omega \tau)\bigr)\,d\omega.
\label{eq:fts}
\end{equation}
Conventionally, the spectrum $S(\omega)$ is then retrieved by an inverse Fourier transform of the fluences sampled on a uniform grid, as prescribed by the Nyquist-Shannon sampling theorem \cite{shannon_communication_1949}. This prescription is so well established that it is rarely questioned -- yet it is worth asking where it comes from, and whether it is the most one can do.

Measuring any continuous-domain quantity fundamentally constitutes a signal reconstruction problem that involves deciding how and where to collect data. The guiding principle for these decisions, at least in the one-dimensional (``time-valued'') case, has long been the Nyquist-Shannon sampling theorem, which sets fundamental sampling constraints for sufficient reconstruction of bandlimited functions in signal processing or spectroscopy \cite{shannon_communication_1949}. However, the emergence of sub-Nyquist sampling methods \cite{mishali2011sub}, most prominently compressed sensing \cite{candes_stable_2006}, suggests that Nyquist-Shannon theory -- while broadly applicable -- is not the most general possible framework. Specifically, these newer approaches highlight that fewer samples can suffice if prior information about the signal structure is known or assumed \cite{candes_stable_2006}.

We can formulate this as an adaptive, decision theoretic view on measurement; one in which the observer actively seeks to maximize information about the system under investigation. From this perspective, determining an \textit{optimal} measurement strategy equates precisely to identifying actions that yield the greatest \textit{information gain}: as each new datum arrives, an estimate of the signal is updated, and one can explicitly calculate how much information each possible measurement is expected to provide, assigning a quantifiable information gain to every potential measurement action.

As we discuss rigorously in the supplemental material, Nyquist sampling emerges in this framework not merely as sufficient but as uniquely optimal when we have no knowledge about the signal under consideration apart from it being constrained by both finite observation duration $T$ and bandwidth $B$. This dual constraint defines a finite informational ``area'', picture a rectangle bounded horizontally by time and vertically by frequency. Within it, the Heisenberg-Gabor uncertainty principle rigorously defines minimal distinguishable information cells -- Gabor's \textit{``logons''} \cite{gabor1946theory} -- with smallest possible area $1/2$. These cells represent independent pieces of information about the signal. Intuitively -- accurate up to a small refinement discussed in the supplemental material -- the total number of these distinguishable cells, the signal's degrees of freedom, is then $2BT$.

Crucially, the observer's initial state of maximal ignorance corresponds mathematically to assigning equal prior value to every information cell: each is equally likely to contain significant signal energy. Under this condition, the rational measurement strategy should distribute samples such that each cell is measured exactly once, without creating information gaps (unmeasured cells) or redundancies (multiple measurements of the same cell). Translating this intuition rigorously into information-theoretic terms, we find that maximizing information gain reduces in this scenario to maximizing the determinant of the Fisher information matrix; a criterion recognized in experimental design theory as ``D-optimality''. Classical results from this theory show that reconstruction within a trigonometric model is optimized when samples are spaced evenly \cite{pukelsheim_optimal_2006}. For our $2BT$ degrees of freedom this then requires a spacing of $1/(2B)$: exactly the Nyquist rate. Hence, Nyquist sampling is not just a minimal criterion for signal reconstruction, but is the uniquely optimal information-acquisition strategy under conditions of maximal uncertainty. In turn, when prior knowledge becomes available, meaning the content of some information cells is partially known while others are suspected to contain more novel information, the optimal measurement spacing adjusts away from uniformity.

Indeed, one seldom measures entirely unknown samples and thus, in virtually every use case \textit{some} prior information on the spectrum is known -- ranging from physical constraints such as the illumination profile and detector response to structural expectations about absorption peaks, bands, and overall smoothness. Conventionally, this knowledge is ignored at the acquisition stage and partially reintroduced \textit{a posteriori} using heuristic data processing such as apodization, smoothing, baseline correction, or peak fitting \cite{baker_using_2014, morais_tutorial_2020}. 

Here, we invert this reasoning and instead adopt a sequential Bayesian inference scheme that uses available knowledge \textit{ab initio} during the measurement, adaptively choosing the delays that yield the greatest information gain relative to the prior. While this requires more complex computations than a fast Fourier transformation, highly optimized hardware and algorithms for matrix operations -- the same techniques powering modern AI systems -- now make these more sophisticated mathematical workflows feasible in real-time. This mirrors the historical context in which FTS itself emerged, becoming broadly adopted only in the 1970s once affordable computing power enabled rapid Fourier transforms \cite{BeckerFarrar1972}. In this sense, our proposed method -- \textit{Active Fourier Spectroscopy} (AFS) -- represents a timely evolution of spectroscopic measurement, capitalizing on vastly expanded computational resources to implement more generally information-optimal measurement strategies.

\section{Active Fourier Spectroscopy}
Technically, AFS can be understood as a sequential Bayesian treatment of the FTS measurement. It works by first encoding our initial belief in the spectrum \( S(\omega) \) as a probability distribution, characterized by a mean \(\mu\) over the spectrum on a discrete frequency grid and a covariance \(\Sigma\) that quantifies the remaining uncertainty and the correlations across frequencies. Each measurement taken at a particular delay \(\tau\) can be viewed as a linear functional of the spectral vector \(S(\omega)\). Discretizing the frequency domain, the relationship between the spectral estimate \(\mu_{\text{prior}}\) and the measured quantity \(F\) at delay \(\tau\) can be expressed through a known linear response operator \(R\) that emerges naturally from Eq.~\ref{eq:fts}. Given \(F\) and \(R\), and assuming Gaussian measurement noise, the updated estimate of the spectrum after incorporating the new observation is:
\begin{equation}
    \begin{aligned}
    \mu_{\text{posterior}} &= \mu_{\text{prior}} + \gamma\bigl(F - R \mu_{\text{prior}}\bigr)\\
    \Sigma_{\text{posterior}} &= (I - \gamma R)\Sigma_{\text{prior}},
    \end{aligned}
    \label{eq:inference}
\end{equation}
where \(\gamma\) is an update weight (see Materials and Methods) that weighs prior belief against measurement noise. Rather than passively assembling the entire interferogram before inverting it, AFS continually refines the spectrum as each measurement arrives, merging existing information with new data in a principled manner.

This adaptive, incremental updating process can be guided by information theory to choose each measurement such that it maximizes the expected reduction in uncertainty \cite{lindley1956measure, fedorov1972theory}. Because the estimate explicitly tracks how uncertain it remains in each region of the spectrum, we can write the expected information gain from a single measurement in closed form as:
\begin{equation}
    \text{Information Gain} = \tfrac{1}{2}\log\frac{|\Sigma_{\text{prior}}|}{|\Sigma_{\text{posterior}}|}.
    \label{eq:information_gain}
\end{equation}
This means that rather than taking measurements at fixed intervals, AFS can select delays that optimally reduce uncertainty. As discussed in the introduction, under total ignorance this objective reduces to uniform Nyquist sampling, which we consider as our baseline case; with prior information the optimal delays redistribute toward regions where the prior is least constraining.

When prior knowledge is available, adaptive measurement faces a critical challenge: how to leverage this knowledge when it is correct, but avoid pathologies that arise when the model assumptions are wrong \cite{sloman_bayesian_2023,sloman_characterizing_2022,go_robust_2022}. To address this, we evaluate the predictive ability of different candidate models -- for example, individualized priors derived from different patient cohorts in a clinical study -- by scoring each according to how well it predicted the running set of measurements \(\mathcal{D} = \{F(\tau_1), F(\tau_2), \ldots\}\) before they were taken, a quantity known as the model evidence. This naturally penalizes overly complex models that must spread their probability mass over many possible outcomes. Models with higher evidence are weighted more strongly, while models that perform poorly relative to the uninformed baseline trigger an automatic fallback to uniform sampling, recovering the exact Nyquist protocol. This mechanism resolves the central tension of using prior information: it delivers substantial gains when the prior is meaningful, while guaranteeing that performance never falls below the conventional FTS baseline if it is not.

A distinctive feature of this formulation is that it introduces no free parameters describing the physical system itself. The effective dimension of the problem is fixed by the time-bandwidth product, and the basis in which the spectrum is represented is fixed by the interferometric measurement of Eq.~\ref{eq:fts}; both follow from physical constraints rather than from a modeling decision. No structural model selection is required because this representation is complete for the measurement at hand -- a completeness confirmed by the emergence of Nyquist sampling under maximal ignorance, which is a consequence of the formulation rather than an assumption fed into it. What remains is the choice of prior beliefs about the sample -- assumptions about spectral smoothness or expected features, encoded through the covariance structure -- and even these are adjudicated by evidence, allowing the data to determine appropriate regularization rather than requiring the practitioner to commit a priori.

Finally, the fully integrated analytical pipeline -- from sequential updates and multi-model weighting to active (optimal) sampling -- can be executed in real time on modern hardware. This analytical tractability is a consequence of formulating spectral inference as a linear-Gaussian model, the same model structure implicit in conventional FTS, of which Nyquist sampling is the maximally-uninformed limit. AFS therefore does not introduce a new approximation but rather makes the existing one explicit and improvable. (This also situates the method within the broader language of optimal experimental design, although here the construction follows entirely from the physics of the measurement; see Supplemental Material, Section~\ref{sec:related_work}.) The Gaussian noise model is appropriate for detector-noise-dominated spectroscopy, the regime of our demonstrated applications (FTIR with thermal detectors, wavefront sensing and imaging with CCD arrays). Extension to photon-limited regimes, where Poisson statistics dominate, requires Monte Carlo sampling for posterior updates and information gain estimation, as demonstrated for parametric spectral models in X-ray photoelectron spectroscopy \cite{nabikaBayesian2024}.

\section{Demonstrations}
\subsection{Molecular fingerprinting}

\begin{figure*}
	\centering
	\includegraphics{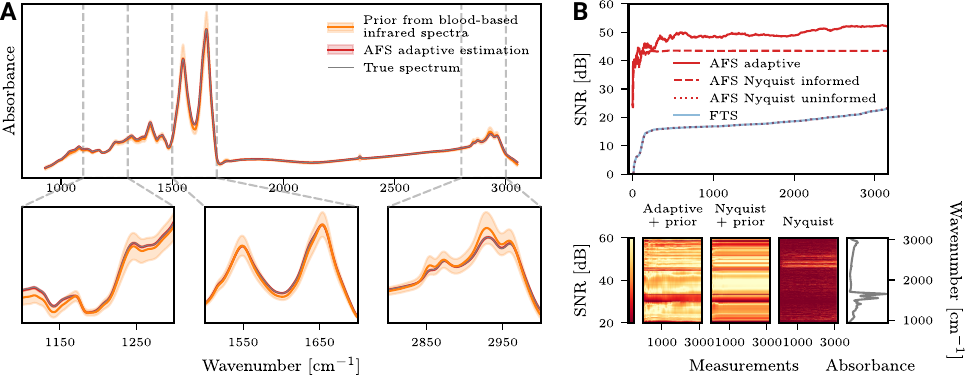}
	\caption{\textbf{AFS reconstruction of a biological FTIR absorbance spectrum.} (\textbf{A}) The prior distribution (orange band) is derived from a set of blood-based infrared spectra exhibits small variance, indicating strong constraints on possible spectra. AFS adaptive estimation (red) closely matches the true spectrum (gray) across the full wavenumber range. Zoom panels highlight key spectral regions showing detailed agreement between reconstruction and ground truth. (\textbf{B}) SNR evolution with measurement count shows AFS adaptive sampling (solid red) reaching a high SNR fast and continuously refining the estimate, followed by AFS with prior knowledge and fixed Nyquist sampling (dashed red) lacking this refinement, while uninformed reconstruction (dotted) performs significantly worse. Wavenumber-resolved SNR maps reveal sampling dynamics, with adaptive AFS exploring spectral regions most efficiently. Right panel shows the reconstructed absorbance spectrum for reference.}
	\label{fig:ftir_absorbance}
\end{figure*}

One of the most prominent applications of FTS is in the medical domain \cite{movasaghi2008fourier, talari2017advances}, with one area of increasing interest being metabolic fingerprinting for disease diagnosis \cite{ellis2006metabolic}.
To demonstrate our approach in this context, we use synthetic data modeled after real infrared measurements of human blood plasma in the framework of the Munich-based Lasers4Life (L4L) clinical study (see Materials and Methods) \cite{huber_infrared_2021}.
Based on these data, previous studies have demonstrated that molecular fingerprinting through vibrational signatures enables the non-invasive detection of cancer using blood plasma samples \cite{kepesidis2025assessing}.

Figure \ref{fig:ftir_absorbance} demonstrates AFS reconstruction of a biological FTIR absorbance spectrum.
The prior distribution, derived from patient data, exhibits small variance in regions with strong biochemical constraints, indicating substantial pre-existing knowledge about possible spectral features.
Building on this informative prior, the posterior estimation achieves even higher precision, evidenced by an uncertainty band too narrow to be visible in the plot.
The reconstruction shows exceptional agreement with the true spectrum across the full wavenumber range ($1000-\SI{3000}{\per\cm}$), with precise reconstruction in key diagnostic regions highlighted in the zoom panels.

The convergence behavior is analyzed in Fig.~\ref{fig:ftir_absorbance} through signal-to-noise ratio (SNR) calculations, defined as the ratio between signal power and reconstruction error power.
AFS demonstrates superior performance in two configurations: adaptive sampling achieves the highest SNR and fastest convergence, while informed Nyquist sampling also shows significant improvement over traditional methods.
We observe that repeated measurements at specific positions can yield more information than the same amount of equally-distributed measurements.
Note that the uninformed Nyquist case performs equivalently to standard FTS, confirming our theoretical prediction of FTS as a limiting case of AFS.

These results demonstrate how AFS can enhance FTIR workflows by incorporating prior knowledge and adaptive sampling strategies while maintaining guaranteed baseline performance through its theoretical connection to traditional FTS.

\subsection{Optical vortices}

\begin{figure}
	\centering
	\includegraphics{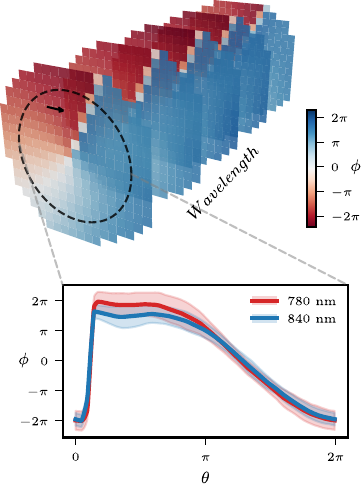}
	\caption{\textbf{Wavelength-resolved phase reconstruction of an optical vortex beam.} Three-dimensional visualization of the beam's phase structure across wavelengths ($750-\SI{850}{\nm}$), showing the characteristic azimuthal phase variation. The zoom panel shows the azimuthal phase profiles extracted at \SI{780}{\nm} (design wavelength) and \SI{840}{\nm} with uncertainty bands derived from covariance propagation through the full AFS spectral reconstruction and phase retrieval chain. The observed wavelength dependence reflects the chromatic behavior of the spiral phase plate.}
	\label{fig:vortex}
\end{figure}

One core strength of Bayesian methods is their rigorous treatment of uncertainty. This is especially important for measurement techniques where the desired quantities are not observed directly but reconstructed from indirect or raw data. Such reconstructions risk compounding errors if uncertainties are not fully analyzed or accounted for, potentially misleading subsequent decisions. A complete uncertainty analysis should therefore also include correlations among reconstructed quantities, represented explicitly through a covariance matrix.

Space-time characterization of ultrafast laser pulses is an example for a complex system where direct measurements of all the relevant quantities are not feasible \cite{alonso_spacetime_2024}.
Of particular interest are optical vortices \cite{heckenberg1992generation} with spectrally-dependent orbital angular momentum \cite{yao2011orbital}.
This type of structured beam has applications in various fields, from relativistic light-matter interactions in the high-intensity regime \cite{vieira_amplification_2016} to quantum computing \cite{fickler_quantum_2014, parigi_storage_2015}.
Wavelength-dependent distortions that affect beam propagation invoke techniques to spectrally resolve the phase structure. Specifically, the ultrafast metrology community has been using various incarnations of imaging FTS to spectrally resolve wavefronts \cite{pariente2016space, piccardo2023broadband, jeandet2020spatio, weisse2023b}.
But in this configuration there exists no principled propagation of correlated uncertainties from their respective spectral reconstructions to quantities calculated further down the analysis chain such as the spectrally-resolved wavefront.

Accurate uncertainty quantification in spectral reconstructions is critically important because the resulting data serve as inputs for detailed physical simulations that underpin our fundamental understanding of these systems. Without uncertainty estimates one cannot assure that the accurate parameter regime is modeled, while overestimated uncertainties would inflate the computational costs of already expensive simulations. A rigorous, balanced uncertainty treatment thus ensures simulations are both reliable and efficient.
We address this issue by employing AFS to spectrally resolve measurements from a wavefront sensor, which not only requires less samples (as demonstrated in the previous section), but allows for uncertainty quantification and propagation.

Our experimental setup combines scanning interferometry with Shack-Hartmann wavefront sensing.
As described in the Materials and Methods section, microlens images of a vortex and reference beam are recorded on a polarization-sensitive sensor at the information-optimal delays to calculate a spectrally resolved estimate which was the starting point for phase reconstruction.
Thus, this method constitutes a full vector beam measurement up to a spectral phase ambiguity.
The AFS reconstruction yields a full spectral covariance matrix for each spatial point, capturing correlations between different wavelength channels.
These correlations are crucial for accurate uncertainty propagation through the phase retrieval pipeline.
The analytical propagation of the covariance through the centroid calculation and phase reconstruction is computationally efficient, requiring only matrix operations scaling with the number of spectral channels.
This approach avoids the computational cost of sampling-based methods while maintaining higher accuracy than simplified diagonal-covariance approximations.
The latter would treat spectral channels as independent, leading to overestimated uncertainties in derived quantities.
Our method thus provides a complete characterization of structured light beams with rigorous uncertainty quantification at manageable computational cost -- see Fig.~\ref{fig:vortex} for a visualization of the results.

\subsection{Hyperspectral imaging}

\begin{figure*}
	\centering
	\includegraphics{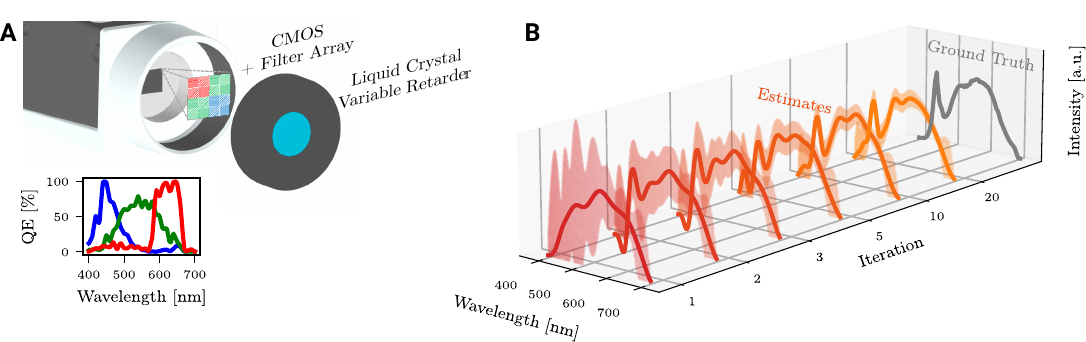}
	\caption{\textbf{Spectral resolution enhancement of an RGB camera using a liquid crystal retarder.} (\textbf{A}) Device overview, consisting of a liquid crystal retarder for delay control between two polarization components and an RGB camera with a colour and polarization filter pixel array, which constitute a compact common-path interferometer with three measurement channels. The inset shows the quantum efficiency of the colour filters. (\textbf{B}) Evolution of the AFS estimate with uncertainty bands as a function of iteration, showing progressive refinement which already improves upon the three-channel resolution after only few measurements.}
	\label{fig:rgb}
\end{figure*}

So far we have considered use cases in which AFS enhanced (imaging) Fourier transform spectroscopy by incorporating prior knowledge and rigorous uncertainty propagation.
Next, we illustrate a new capability emergent from our more general framework: the seamless fusion of multiple, partial information streams.
The example we consider is hyperspectral imaging, a technology used for spatio-spectral discrimination across diverse domains from medical imaging \cite{lu2014medical}, food safety and agriculture \cite{feng_application_2012, dale_hyperspectral_2013}, materials science \cite{dong_review_2019} to remote sensing \cite{goetz_three_2009}.
Common implementations require either extensive spatial or spectral scanning or complex geometric arrangements \cite{hagen_review_2013} and thus are subject to fundamental trade-offs of spatial and spectral resolution, measurement speed and computational resources \cite{arce_compressive_2014}.

In our proof-of-concept demonstration, we realize a compact hyperspectral imager by combining coarse spectral information from color filters in an RGB camera with strategically chosen interferometric delay.
Our setup uses a liquid crystal retarder as common-path interferometer immediately in front of the sensor. The two delayed components are superimposed on polarization-filtered pixels of an RGB camera to yield the autocorrelation measurements.
For each delay step, the camera captures an image where different pixel groups -- corresponding to the red, green, and blue filters -- encode spectral information through a combination of their inherent filter transmission curves and interference patterns modulated according to Eq.~\ref{eq:fts}.
While the color filters alone provide only a coarse spectral resolution, the additional time-dependent modulations encode finer spectral details.
Building on the basic AFS inference algorithm Eq.~\ref{eq:inference}, we then integrate the known spectral responses of the RGB filters with the measured interferometric autocorrelation curves.
By iteratively updating the spectral estimate, AFS systematically combines the limited but complementary spectral information from the different color channels and delay positions.
The algorithm selects optimal delay steps to maximize information gain, ensuring rapid and precise spectral reconstruction with minimal measurements.

As shown in Fig.~\ref{fig:rgb}, the reconstructed spectra rapidly converge to high fidelity within just a few strategically chosen delay measurements.
Thereby this approach effectively transforms the RGB camera into a compact hyperspectral imager, capable of resolving detailed spectral features beyond the native resolution of its color filters while maintaining full spatial resolution.
It should be noted that the present measurements included minimal prior information.
Contextual priors derived from scene information or meta-data can effectively complement this, providing a fertile ground for developing fast and compact hyperspectral imagers.

\section{Discussion}
We have introduced a framework that operationalizes adaptive measurements among different variations of Fourier-based spectroscopy, based on theoretical work that connects classical sampling theory as special case of Bayesian decision making when operating with uninformed priors. This conceptual bridge enables substantially accelerated measurements when prior knowledge exists. Analytical work guarantees performance at least matching conventional methods -- meaning that our approach can \textit{upgrade} many existing setups to achieve both faster and more reliable, uncertainty quantified results. 

Our demonstrations across three distinct domains of optical spectroscopy -- clinical diagnostics, optical vector-field characterization, and hyperspectral imaging -- reveal the method's versatility and immediate practical impact. In each case, we observe accelerated measurements and can provide rigorous uncertainty bounds with fully propagated covariances, enabling confident decision-making in scenarios where conventional methods offer only point estimates or simplified error models.
This is particularly valuable in data-driven science and medicine, which is increasingly relying on automated processing pipelines and decision systems. Here, our rigorous uncertainty quantification addresses a critical gap between raw measurements and evidence-based decision-making. We also demonstrated the unique potential for new diagnostics that merge information from different data streams, showcased by combining RGB sensors with autocorrelation measurements to realize hyperspectral sensors. This optimally combines the strengths of different measurement modalities into new hybrid measurement schemes.

Our work naturally extends in several directions: The current implementation employs multivariate Gaussian priors, but the mathematical framework supports more sophisticated constructions. Future implementations could incorporate Bayesian neural networks \cite{jospin2022hands} to capture complex spectral relationships conditioned on relevant metadata, e.g. individualized or group-specific priors for precision health care. Similarly, more direct data streams could be integrated, e.g. combining many spectral channels with suitable delay measurements. Furthermore, depending on the application the optimization goal itself may be adapted from pure information gain to more objective-driven information metrics, such as maximum-value entropy search \cite{wang2017max}. Beyond specific extensions, our framework exemplifies a fundamental shift in measurement science: from static protocols to dynamic, knowledge-driven strategies.
This "agentic" approach recognizes that in many scenarios, measurement is not about collecting predetermined data points but about extracting the most informative ones given our prior knowledge.
This transparent, bottom-up approach starting from available knowledge complements deep learning approaches.

As computational capabilities continue to advance, this approach paves the way for a new generation of scientific instruments that accumulate knowledge in a truly optimal way. In doing so, our work addresses a blind spot in the rapidly growing literature on autonomous experiments, self-driving laboratories, and agentic scientific instruments such as \cite{tobias2025autonomous,wang2026agentoptics,yang2026qiushi,uddin2025optics,sha2025qcopilot}. 
Much of this recent work realizes autonomy in a top-down manner: a deep-learning model or LLM agent is placed above the experiment and tasked to actively coordinate. 
Such systems are powerful, but their decisions depend on learned representations and training distributions whose latent biases generally cannot easily be audited nor allow for disciplined continual operation. AFS realizes a different type of agency that is bottom-up and purely decision-theoretic: the possible actions are fixed by the physics of Fourier spectroscopy, the belief state is an explicit posterior over the spectrum, and every measurement is chosen by a transparent information-gain objective. The resulting autonomy is therefore not achieved via an opaque intelligence acting on the instrument, but an interpretable statistical decision rule embedded in the measurement process itself. This distinction is central for the applications in which FTS is most valuable. 
In science, medicine, metrology, and diagnostics, faster acquisition alone is not sufficient if the resulting measurement record cannot be trusted, audited, or assigned meaningful uncertainty. 
AFS provides guarantees of a kind that compressed sensing, deep learning, and LLM-agent control do not generally provide. It therefore contributes an interpretable and reliability-oriented form of agentic measurement: one that actively exploits prior knowledge to improve outcomes, while maintaining the trustworthiness that make FTS a standard tool in scientific and medical settings.

\bibliographystyle{apsrev4-2}

\begin{acknowledgments}
We thank Prof. Ferenc Krausz for supporting this work. We thank the Federal Republic of Germany and the Free State of Bavaria for funding the CALA infrastructure (15171 E 0002) and its operation. J.S. and J.E. thank the International Max Planck Research School for Advanced Photon Science (IMPRS-APS) for support. 

This work was supported by the Independent Junior Research Group "Characterization and control of high-intensity laser pulses for particle acceleration", DFG Project No.~453619281. C.E. is funded by BMBF through the MACLIP project.

A.D. and J.S. developed the mathematical framework. J.S. performed and analyzed experiments with the help of S.H., C.E. and J.E. The L4L dataset was provided by N.L.-K. and K.V.K. A.D. and J.S. wrote the manuscript with input from all co-authors. A.D. supervised the work.

The datasets generated during and/or analyzed during the current study and the code implementing Active Fourier Spectroscopy are available from the corresponding author (J.S.) upon reasonable request. 

The authors declare no competing interests.

\end{acknowledgments}

\clearpage
\appendix
\onecolumngrid

\section*{Supplementary Materials}
\setcounter{tocdepth}{3}  

\setcounter{secnumdepth}{4}  
\renewcommand{\thesection}{S\arabic{section}}
\renewcommand{\thesubsection}{S\arabic{section}.\arabic{subsection}}
\renewcommand{\thesubsubsection}{S\arabic{section}.\arabic{subsection}.\arabic{subsubsection}}
\renewcommand{\theparagraph}{S\arabic{section}.\arabic{subsection}.\arabic{subsubsection}.\arabic{paragraph}}
\makeatletter
\renewcommand{\p@subsection}{}
\renewcommand{\p@subsubsection}{}
\renewcommand{\p@paragraph}{}
\makeatother
\setcounter{section}{0}  

\section{Materials and Methods}\label{sec:methods}
\subsection{General AFS inference procedure}
\label{sec:bas_procedure}

AFS combines sequential Bayesian inference with information-optimal design selection. The method maintains one or more candidate spectral models, each represented by a Gaussian distribution over the discretized spectrum, and adaptively selects measurement delays to maximize expected information gain. Algorithm~\ref{alg:bas} summarizes the complete procedure; we briefly describe each component here, with detailed derivations in subsequent sections.

The spectrum is discretized on a frequency grid with spacing determined by the observation window and bandwidth. Each measurement at delay $\tau$ yields a scalar fluence related to the spectrum through the linear model $F = R(\tau) S + \varepsilon$ with $\varepsilon \sim \mathcal{N}(0, \sigma_F^2)$, where $R(\tau)$ encodes the interferometric response.

For each candidate prior $k$ with moments $(\mu_k, \Sigma_k)$, the posterior update follows standard Bayesian linear-Gaussian inference:
\begin{align}
    \mu_k' &= \mu_k + \gamma_k (F - R(\tau)\mu_k), \qquad
    \Sigma_k' = (I - \gamma_k R(\tau))\Sigma_k,
\end{align}
where $\gamma_k = \Sigma_k R(\tau)^\top (R(\tau)\Sigma_k R(\tau)^\top + \sigma_F^2)^{-1}$ is the Bayesian update weight (see Supplemental Material, Section~\ref{sec:linear_fts}). Model weights are updated based on predictive performance via Bayesian evidence (see Supplemental Material, Section~\ref{section:prior_selection}).

Design selection maximizes the expected information gain, computed as a weighted average across models: $\mathrm{IG}(\tau) = \sum_k w_k \, \mathrm{IG}_k(\tau)$, where each term has the closed form $\mathrm{IG}_k(\tau) = \frac{1}{2}\log(1 + R(\tau)\Sigma_k R(\tau)^\top / \sigma_F^2)$. This approximation is analyzed in the Supplemental Material, Section~\ref{sec:mixture_ig}.

\begin{table}[h]
\centering
\fbox{\parbox{0.95\linewidth}{
\textbf{Algorithm: Active Fourier Spectroscopy (AFS)}
\vspace{0.5em}
\hrule
\vspace{0.5em}
\textbf{Input:} Candidate priors $\{(\mu_k^{(0)}, \Sigma_k^{(0)})\}_{k=1}^K$, noise variance $\sigma_F^2$, candidate delays $\mathcal{T}$, stopping criterion

\textbf{Initialize:} $n \gets 0$; $w_k \gets 1/K$; $L_k \gets 0$ for all $k$

\textbf{Repeat:}
\begin{enumerate}
    \item \textbf{Design:} Compute $\mathrm{IG}(\tau) = \sum_k w_k \, \mathrm{IG}_k(\tau)$ for all $\tau \in \mathcal{T}$; select $\tau^* = \arg\max_\tau \mathrm{IG}(\tau)$
    \item \textbf{Measure:} Acquire fluence $F$ at delay $\tau^*$; increment $n \gets n+1$
    \item \textbf{Update posteriors:} For each model $k$, compute $\gamma_k$ and update $\mu_k, \Sigma_k$
    \item \textbf{Update weights:} Accumulate log-evidence $L_k \gets L_k + \log p(F|\tau^*, k)$; recompute $w_k \propto \exp(L_k/T)$
\end{enumerate}
\textbf{Until} stopping criterion met (e.g., $n = N_{\max}$ or uncertainty threshold reached)

\textbf{Output:} Posterior means $\mu_k$, covariances $\Sigma_k$, weights $w_k$
}}
\caption{Complete AFS procedure for adaptive spectral reconstruction.}
\label{alg:bas}
\end{table}

\subsection{Synthetic data on molecular fingerprinting}

Infrared measurements of liquid biopsies are typically performed on commercial FTIR devices, which do not provide direct access to time-domain data.
Our analysis of molecular fingerprinting was conducted using synthetic spectra modeled after blood-based spectra measured via FTIR, collected as part of the L4L study -- a clinical trial at Ludwig-Maximilians-Universit{\"{a}}t M\"unchen, registered under ID DRKS00013217 in the German Clinical Trials Register (DRKS).
These data mirror the experimental methodology of Huber et al. \cite{huber_infrared_2021, kepesidis2025assessing}.
The study applied FTIR spectroscopy to blood-based liquid biopsies, where molecular fingerprinting through vibrational signatures enables non-invasive detection of cancer markers in blood plasma.

To allow working with de-identified spectra, a statistically matched synthetic set was generated. This synthetic dataset was constructed by fitting multivariate Gaussian models separately to spectra from lung cancer patients and healthy controls, which is consistent with approaches previously used in spectroscopic data simulation \cite{beleites_sample_2013, eissa_limits_2023}.
The models capture group-specific mean spectra and covariance matrices, with statistical validation through comparison of spectral distributions, covariance patterns, and principal component analysis confirming faithful reproduction of the biochemical characteristics observed in clinical samples.
Furthermore, infrared measurements of quality control (QC) samples with no biological variation provide a means to assess the reproducibility of the measurements.

The demonstration used synthetic FTIR data modeling biological tissue absorption.
Absorbance spectra were converted to intensity spectra assuming a $T=\SI{1200}{\K}$ blackbody illumination.
The dataset was split into training and test sets, with training set means and covariances serving as prior models.
Time-domain measurements were simulated using random test set samples with time-domain noise that was inferred from the QC samples.
Performance evaluation converted reconstructed spectra back to absorbance.

\subsection{Vortex measurements}
A Ti:Sapphire oscillator output ($750-\SI{850}{\nm}$) was passed through a quarter-wave plate and spiral phase plate (design charge $\ell$=2 at $\SI{780}{\nm}$) to generate the vector beam.
The beam entered a computer-controlled Michelson interferometer, with the output imaged onto a microlens array (focal length \SI{5.2}{\mm}, pitch \SI{150}{\um}) focusing onto a polarization-sensitive CMOS camera (Sony IMX250MZR, $2448\times2048$ pixels).
AFS reconstruction was performed independently for each microlens focus point, yielding a hyperspectral datacube and a spectral covariance, starting from a prior set with zero mean and RBF covariances to enforce the minimal assumption of spectral smoothness.
Phase gradients were extracted by calculating focal spot centroid displacements between the vortex measurement and a flat reference beam, using regions of interest around each microlens focus.
Ideal vortex derivatives were fit to the phase gradients with a subsequent zonal reconstruction of the residual and its covariance.

\subsection{RGB camera measurements}
We demonstrated AFS's ability to enhance spectral resolution of RGB cameras using a basic optical setup: Light from a fiber-coupled light source was collimated, polarized using a polarization filter oriented to produce equal transversal contributions on the extra- and ordinary axes of the subsequent liquid crystal variable retarder before detection on a color CMOS sensor with polarization filters (Sony IMX250MYR).
This yielded normalized measurements in the three colour channels $c$:
\[
F_c(\tau) = F^{45^{\circ}}(\tau) / \left(F^{0^{\circ}}(\tau) + F^{90^{\circ}}(\tau)\right)    
\]
The liquid crystal device provided voltage-controlled phase delays, calibrated using monochromatic illumination.
AFS reconstruction incorporated the known RGB filter curves and polarization response to recover high-resolution spectra from the combined color and delay measurements, starting from a prior set with zero mean and RBF covariances to enforce the minimal assumption of spectral smoothness.

\section{Related Work and Contribution}\label{sec:related_work}

Recent computational advances have dramatically expanded the practical applicability of Bayesian experimental design, with successful demonstrations across diverse scientific domains. This progress provides the foundation for addressing more fundamental questions about the nature of optimal measurement itself.

\subsection{Inference}

Inference deals with reasoning about any latent phenomena given some observations. Bayesian inference (BI) in particular relies on Bayes' theorem. This enables the incorporation of prior knowledge in regression problems and provides well-motivated uncertainty estimates from limited data without resorting to asymptotic limits. The latent variables can range from collections of parameters in known models to entire functions.

Due to sampling costs, many applications involving Bayesian inference are restricted to an offline setting. An important exception are linear Gaussian models, which allow for analytical closed-form updates. These are classically associated with Kalman filtering and smoothing in state-space form, but the same conjugate structure also appears in Bayesian linear regression (see also Sec.~\ref{sec:linear_inference}), Wiener filtering, Gaussian-process regression, and linear inverse problems. 

For the particular case of Fourier transform spectroscopy the inference problem is usually formulated in a Frequentist formalism. Bayesian inference improves the conditioning of the inference problem by adding prior knowledge. 

Apart from the present work, Schmuck et al. applied Bayesian inference to analyze fixed, uniformly sampled datasets from Fourier spectroscopy \cite{schmuckFourier2017a}. They select a Brownian-bridge kernel that enforces vanishing spectra at band edges, yielding closed-form posteriors with a non-stationary variance profile that peaks in the middle of the analysis band. Their approach addresses mathematical well-posedness for the inversion problem under uniform sampling.

The bridge between inference and Shannon's information theory that is used throughout our paper was developed by Lindley in the 1950s\cite{lindley1956measure}. Lindley proposed to quantify the informativeness of an observation or experiment by the expected reduction in entropy when updating a prior to a posterior. His motivation was epistemic: to compare experiments in terms of the knowledge they provide about the parameters of interest, even in the absence of any specific decision problem. 

\subsection{Optimal Experimental Design}

Classical optimal design formalizes how informative a design is through functionals of the (expected) Fisher information matrix. Let $\xi$ denote a design (a probability measure over candidate settings $x\in\mathcal{X}$ with weights $\{w_i\}$), and let
\begin{equation}
    \mathcal{I}(\theta;\xi)
    \;=\;
    \mathbb{E}_{y\,|\,\theta,\xi}\!\big[ -\nabla_\theta^2 \log p(y\mid \theta,\xi)\big]
\end{equation}
be the information matrix under model parameter $\theta$. In linear-Gaussian regression with basis $f(x)$ and noise covariance~$\Sigma_\varepsilon$, this reduces to
$\mathcal{I}(\theta;\xi)=\sum_i w_i\, f(x_i)^\top \Sigma_\varepsilon^{-1} f(x_i)$ (independent of $\theta$), which makes the role of the measurement settings explicit (see also Sec.~\ref{sec:linear_inference}). Alphabet-optimality criteria\cite{fedorov1972theory, pukelsheim_optimal_2006} pick $\xi$ by optimizing a scalar summary of $\mathcal{I}$; common choices are:

\begin{description}
    \item[D-optimality] maximizes the generalized precision:
    \(
        \max_\xi \det \mathcal{I}(\theta;\xi)
        \quad\text{equivalently}\quad
        \min_\xi \log\det \mathcal{I}(\theta;\xi)^{-1}.
    \)
    This minimizes the \emph{volume} of the confidence ellipsoid and has deep ties to information theory: in regular models, local maximization of $\log\det \mathcal{I}$ coincides with maximizing a Laplace/Gaussian approximation to mutual information.

    \item[E-optimality] maximizes the worst-case (smallest) information direction:
    \(
        \max_\xi \lambda_{\min}\!\big(\mathcal{I}(\theta;\xi)\big),
    \)
    thereby guarding against ill-conditioning by shortening the longest semi-axis of the ellipsoid.
    \item[G-optimality] minimizes the \emph{maximum} prediction variance,
    \(
        \min_\xi \sup_{x\in\mathcal{X}} v(x;\xi).
    \)
\end{description}

In linear models with homoskedastic noise, the Kiefer-Wolfowitz equivalence theorems connect D- and G-optimality\cite{kiefer1974general}, while E-optimality only coincides with the former under specific conditions.

\subsubsection{Bayesian Experimental Design}

As design moved from Frequentist to Bayesian formulations, two strands converged. As stated before, classical optimal design optimized functionals of the Fisher information. In parallel, Bayesian decision theory framed design as maximizing expected utility, and developments in random utility theory made it natural to treat utility as stochastic -- either via outcome randomness or genuine preference uncertainty. Stone\cite{stone1959application} and Fedorov (Ch. 7) \cite{fedorov1972theory} then studied designs that maximize the expected Shannon information (mutual information) as an accuracy criterion. This paralleled Lindley's earlier developments, who had used the same information measure framed as part of inference, not decision. in 1979, Bernardo supplied the decision-theoretic link by noting that maximizing information can be expressed as a special case of maximizing the expected log score of utility \cite{bernardo1979expected}. This integrated and subsumed Lindley's inference work and Federov's information theoretic decision criteria under the umbrella of the emerging field of Bayesian experimental design \cite{chaloner1995bayesian}. Based on its academic lineage, BED may best be described as the problem to \textit{improve the expected utility over a finite set of experimental parameters given the available prior knowledge}. 

Modern theoretical advances have addressed computational bottlenecks arising from intractable probability densities and expectations thereof \cite{rainforthModern2023}. This essentially leads to optimization steps within the inference loop, which itself constitutes another optimization problem. A large portion of BED concerns the offline optimization of experiments, while the less common sequential implementation of BED within an experiment is sometimes referred to as Bayesian adaptive design (BAD). As discussed by Rainforth and colleagues, \textit{"BED, and in particular BAD, can suffer from serious pathologies in practice if our model is misspecified"} \cite{rainforthModern2023}.

Despite these limitations, Bayesian adaptive design has proven highly efficient when optimizing experimental settings to estimate selected few parameters of physical models -- typically 2-5 parameters such as decay rates or frequencies: Quantum sensing applications have demonstrated substantial efficiency gains through adaptive inference of system parameters including Hamiltonian coefficients, resonance frequencies, decay rates, and quantum phases using various measurement modalities from spin-dependent photon detection to interferometric protocols \cite{mcmichaelSequential2021, granadeRobust2012, dushenkoSequential2020a, valeriExperimental2020, caouette-mansourRobust2022, sarraDeep2023, deneeveTimeadaptive2025}. Similarly, astronomical applications have explored both adaptive and non-adaptive optimization approaches for parameter estimation problems, ranging from toy model demonstrations to practical measurement design for cosmological parameter inference under observational constraints \cite{loredoBayesian2004, paykariSparsely2013}. One should note that the closely related field of Bayesian active learning -- a machine learning application which is sometimes considered isomorphic to BED\cite{rainforthModern2023} -- routinely features higher dimensional design spaces, e.g. using Bayesian Active Learning by Disagreement\cite{houlsby2011bayesian}.

\subsubsection{Bayesian Optimization}

A related and often conflated application of Bayesian decision making is Bayesian Optimization (BO). Technically, Bayesian optimization is a global optimization method seeking extrema of a function guided by an acquisition functions that weighs expectation against uncertainty. That definition itself could be subsumed under the umbrella of BED as one equates observed function value with a utility measure or, equivalently, one can treat a utility as function value and subsume BED under BO. 
However, in practice the distinction between the fields boils usually down two a question of framing and scope. 

\begin{itemize}
    \item First, BO is usually understood as a sequential Bayesian model-based optimization strategy, with an emphasis on the cheap nature of surrogate models compared to evaluations of the real system. Its origin is widely attributed to geostatistics, where D. G. Krige's introduced surrogate modeling for finding sites for gold mining in the 1950s\cite{krige1951statistical, matheron1963principles}, and sequential decision rules developed by Kushner and Mockus completed these over the following two decades,  to form a recognizable framework BO using probability \cite{kushnerNew1964} or expectation\cite{mockusBayesian1975} of improvement. 
    \item With regards to scope, on can argue for a taxonomy wherein BED concerns finite or low-dimensional parameter-centric design (more often than not offline), while BO deals with online optimization over (infinite) function spaces that are described using non-parametric, kernelized priors. The exception proves the rule and counter examples for both exist. For instance, non-sequential "batch" optimization exists in BO\cite{gonzalez2016batch}, as well as direct modeling of utility through pairwise Gaussian processes\cite{gonzalez2017preferential}.
\end{itemize}

The actual decision rules in BO are encoded in the acquisition functions. These include expectation methods similar to BED, like the aforementioned expected improvement, but also regret bounds such as confidence bounds from the multi-armed bandit literature. The introduction of information-based optimization metrics to the field is a relatively recent development, most prominently known as "entropy search"\cite{hennig2012entropy}. Information gain is for instance expressed with respect to position or value of the extreme value\cite{wang2017max}. But because of the high computational sampling cost needed to compute information metrics, practitioners frequently resort to expected improvement as a default policy.

\subsection{Signal Processing}

Classical signal processing is built around \emph{sufficiency}: sampling prescriptions that guarantee stable reconstruction for an entire model class, regardless of the particular signal instance. The Nyquist--Shannon theorem is the canonical example. It asserts that every function in the bandlimited class $\mathcal{B}_B$ is reconstructible from uniformly spaced samples at rate $2B$. This is a worst-case statement over the whole class -- no prior knowledge selects one spectrum over another -- and it certifies that \emph{no Fourier mode is under-sampled}. In the language of optimal experiment design, this is most naturally read as an \emph{E-optimal} guarantee: a uniform \emph{floor} on the information available in every relevant direction.

Compressed sensing (CS) extends this sufficiency ethos to structured model classes by replacing "all Fourier modes" with "$k$-sparse supports." Measurements $\mathbf{y}=A\mathbf{x}+\boldsymbol{\epsilon}$ with noise covariance $\Sigma$ furnish the Fisher information matrix $J=A^\top\Sigma^{-1}A$, which quantifies the information content of measurements about the unknown signal $\mathbf{x}$.

The Restricted Isometry Property (RIP) is CS's fundamental guarantee: for every $k$-sparse signal (having at most $k$ non-zero components), the eigenvalues of the restricted Fisher matrix $J_{SS}$ must satisfy $\epsilon \le \lambda_{\min}\!\big(J_{SS}\big) \le \lambda_{\max}\!\big(J_{SS}\big) \le \epsilon'$.
The lower bound $\lambda_{\min} \ge \epsilon$ ensures sufficient information to reconstruct any $k$-sparse signal---every sparse configuration remains identifiable from the measurements. The upper bound $\lambda_{\max} \le \epsilon'$ prevents excessive measurement redundancy in any direction, maintaining the conditioning of the inverse problem. Together, these bounds guarantee that all possible $k$-sparse signals are measured with comparable fidelity.

Random measurement matrices (subgaussian, partial Fourier) achieve RIP with high probability precisely because their lack of structure prevents systematic under- or over-sampling of any sparse support. By distributing measurement sensitivity uniformly throughout the signal space, random constructions naturally satisfy the required eigenvalue bounds across all sparse subspaces. This constitutes a sufficiency strategy: rather than optimizing measurements for a particular signal instance, the design ensures adequate performance for the entire signal class.

Viewed through this lens, classical Nyquist sampling and compressed sensing follow the same principle: both enforce minimum information thresholds. Nyquist sampling maintains uniform information across all Fourier modes (equispaced samples weight all frequencies equally), while CS maintains adequate information across all sparse supports (RIP bounds the Fisher matrix eigenvalues uniformly). In both frameworks, the critical constraint is a lower bound on the smallest eigenvalue of the relevant Fisher matrix---guaranteeing that no signal component lacks sufficient measurement information. This E-optimality criterion (maximizing worst-case information) contrasts with D-optimality (maximizing total information via $\log\det J$), which permits redistributing information between different signal components to maximize an aggregate measure.

Our framework uses this distinction constructively: under maximal ignorance the symmetry forces $M\propto I$ and the Nyquist design is simultaneously E- and D-optimal; when informative priors or heteroskedastic responses break symmetry, AFS departs from uniform spacing by optimizing a global information functional.

\begin{table*}[t]
\centering
\small
\renewcommand{\arraystretch}{1.2}
\setlength{\tabcolsep}{4pt}
\begin{tabular}{@{}>{\raggedright\arraybackslash}p{2.7cm}>{\raggedright\arraybackslash}p{3.7cm}>{\raggedright\arraybackslash}p{3.3cm}>{\raggedright\arraybackslash}p{3.2cm}>{\raggedright\arraybackslash}p{3.3cm}@{}}
\hline
\textbf{Aspect} &
\textbf{Information-optimal measurement (this work)} &
\textbf{Bayesian experimental design (BED)} &
\textbf{Bayesian optimization (BO)} &
\textbf{Signal Processing (Nyquist / compressed sensing)} \\
\hline
Optimization goal &
Maximize expected \textit{information gain} (entropy reduction); reduces to D-optimality under ignorance &
Maximize expected \textit{utility} for parameter inference &
Maximize/minimize objective via acquisition (EI/UCB/entropy) &
Guarantee \textit{sufficiency} (E-optimal floor / RIP), not info-optimal \\
Optimized quantity &
\textit{Full function} (e.g. spectrum) with posterior mean \& \textit{full covariance} &
\textit{Finite parameters} of a specified model &
\textit{Extremum} of a function (not full reconstruction) &
Signal in a constrained class (bandlimited/sparse) \\
Design mode &
\textbf{Online} sequential, real-time thanks to analytical solutions; batch optional &
Mostly \textit{offline}; adaptive BAD exists but heavy &
\textbf{Online} sequential by default (batch variants) &
\textit{Offline} fixed designs (uniform/random) \\
Priors \& safety &
Multi-prior, \textit{evidence-weighted}; \textbf{safe fallback} to uninformed \(\Rightarrow\) performance \(\ge\) Nyquist &
Single prior; misspecification can harm; \textit{no safety guarantee} &
Surrogate prior (kernel); can mislead; \textit{no safety guarantee} &
Minimal priors (ignorance/sparsity); no adaptivity \\
Uncertainty output &
\textit{Full posterior covariance} across channels; propagated to derived quantities &
Posterior over parameters; propagation beyond model uncommon &
Surrogate uncertainty for search; rarely propagated to outputs &
Bounds/coherence metrics; no full covariance \\
\hline
\end{tabular}
\caption{Framework comparison indicating \textit{"typical"} regimes within experimental physics for each methodology. Note that taxonomy is fluid and BED can, for instance, be more widely seen as an overarching framework for any of these. Our work is \textit{uniquely positioned} within by its operational combination: it optimizes information for full signal reconstruction, runs online with evidence-weighted priors, and is safe (reverts to Nyquist/FTS under ignorance).}
\label{tab:concise_method_comparison}
\end{table*}

\subsection{Our contribution}

Our work operationalizes Bayesian experimental design for interferometric spectroscopy. Depending on the viewpoint, it can be considered as either a a generalization of classical sampling theory or special case of Bayesian decision making. The most suitable characterization may be a sequential decision-making (policy) viewpoint -- closely related to optimal-control language -- because it matches the ``instrument as an agent'' interpretation: the device maintains a belief state (posterior) and selects the next action (delay) to maximize an information-based utility.

While the former approaches are optimal within their probabilistic formulations, the model structure (parameterization, basis choice) is typically a modeling decision rather than being derived from physical constraints alone.

We reframe sampling-based measurements in spectroscopy as information-optimal reconstruction, where the parameterization emerges from physical constraints (time-bandwidth limitation) rather than being chosen for tractability. As the Fourier basis for interferometric measurements is complete, this guarantees capturing the entire signal rather than a parametric model representation. Time-frequency concentration arguments allow for discrete computational representation of such continuous signals.

This leads to several qualitative advances. We prove that Nyquist sampling emerges from information-optimal sampling under complete ignorance, providing theoretical grounding for classical sampling. Through this result and using Bayesian model evidence, we guarantee that performance never falls below conventional methods. This contrasts with various approaches that recognize the potential suboptimality of Nyquist sampling, such as compressed sensing methods that exploit sparsity assumptions when appropriate bases are known, or heuristic sampling optimization schemes that target specific loss functions. However, as discussed above, these approaches lack performance guarantees and may not improve upon classical methods when their underlying assumptions are violated.

It also contrasts with recent work that has identified robustness challenges when Bayesian experimental design encounters model uncertainty. While approaches like robust expected information gain \cite{go_robust_2022} and explicit nuisance parameter management \cite{sloman_bayesian_2023} address these issues, they require computational overhead or pre-specified parameter hierarchies and, crucially, lack the performance guarantees we establish through our connection to Nyquist sampling. Our empirical Bayes approach provides computational efficiency through analytical updates, automatic model selection via evidence comparison, and safety guarantees through fallback to classical sampling ---- advantages enabled by working in spectroscopy's natural Fourier basis rather than parametric model spaces.

We address high-dimensional signal reconstruction through explicit temporal-to-spectral domain translations, where concentration arguments enable efficient yet complete reconstruction. This extends existing work in adaptive spectral reconstruction \cite{itoOptimal2025} by providing theoretical underpinnings and real-time capabilities that enable practical implementation.

\section{Detailed Derivations and Discussions}
\subsection{Bayesian inference model}

In the following, we formalize the Bayesian inference approach underlying our information-optimal measurement strategy, explicitly linking prior knowledge with measurement actions. We adopt a general linear Gaussian inference framework, widely used for its analytical tractability, which provides a rigorous foundation for adaptive and uncertainty-aware sampling strategies.

\subsubsection{General linear Gaussian inference}\label{sec:linear_inference}

Let us consider a general Bayesian inference problem with the following elements:
\begin{itemize}
    \item An unknown parameter $\theta \in \Theta$, where $\Theta$ may be finite-dimensional or an infinite-dimensional function space
    \item A set of possible measurement actions $x \in \mathcal{X}$
    \item A linear measurement model relating observations to parameters
    \item Gaussian prior and likelihood models
\end{itemize}

For a set of measurement actions $\{x_1, x_2, \ldots, x_n\} \subset \mathcal{X}$, the linear observation model takes the form:
\begin{equation}
    y_i = \mathcal{A}(x_i)\theta + \varepsilon_i, \quad i = 1,2,\ldots,n
\end{equation}
where $\mathcal{A}(x_i)$ is the measurement operator associated with action $x_i$, and $\varepsilon_i \sim \mathcal{N}(0, \Sigma_\varepsilon(x_i))$ represents observation noise that may depend on the measurement action.

In the compact vector form, the measurement model becomes:
\begin{equation}
    \mathbf{y} = \mathbf{A}\theta + \boldsymbol{\varepsilon}
\end{equation}
where $\mathbf{A}$ combines the measurement operators for all actions, and $\boldsymbol{\varepsilon} \sim \mathcal{N}(\mathbf{0}, \mathbf{\Sigma}_\varepsilon)$.

Starting with a Gaussian prior belief about the parameter:
\begin{equation}
    p(\theta) = \mathcal{N}(\mu_0, \Sigma_0)
\end{equation}

We apply Bayes' theorem to obtain the posterior distribution:
\begin{equation}
    p(\theta|\mathbf{y}) \propto p(\mathbf{y}|\theta)p(\theta)
\end{equation}

Due to the conjugacy properties of Gaussian distributions with linear models, the posterior is also Gaussian:
\begin{equation}
    p(\theta|\mathbf{y}) = \mathcal{N}(\mu_{\text{post}}, \Sigma_{\text{post}})
\end{equation}

where the posterior moments can be computed as

\begin{equation}
    \Sigma_{\text{post}} = \Sigma_0 - \Sigma_0\mathbf{A}^T(\mathbf{A}\Sigma_0\mathbf{A}^T + \Sigma_\varepsilon)^{-1}\mathbf{A}\Sigma_0
\end{equation}
\begin{equation}
    \mu_{\text{post}} = \mu_0 + \Sigma_0\mathbf{A}^T(\mathbf{A}\Sigma_0\mathbf{A}^T + \Sigma_\varepsilon)^{-1}(\mathbf{y} - \mathbf{A}\mu_0)
\end{equation}

\subsubsection{Limited rank update through finite measurements}

For sequential updates with individual observations, we can define:
\begin{equation}
    \mathbf{k}(x) = \Sigma_{\text{current}}\mathcal{A}(x)^T
\end{equation}
\begin{equation}
    s(x) = \mathcal{A}(x)\Sigma_{\text{current}}\mathcal{A}(x)^T + \Sigma_\varepsilon(x)
\end{equation}

Then the update for a new measurement at action $x$ becomes:
\begin{equation}
    \mu_{\text{new}} = \mu_{\text{current}} + \mathbf{k}(x)s(x)^{-1}(y - \mathcal{A}(x)\mu_{\text{current}})
\end{equation}
\begin{equation}
    \Sigma_{\text{new}} = \Sigma_{\text{current}} - \mathbf{k}(x)s(x)^{-1}\mathbf{k}(x)^T
\end{equation}

A fundamental property of these updates is that they have limited rank, regardless of the dimensionality of the parameter space $\Theta$. To see this, examine the difference between prior and posterior covariances for sequential measurements

\begin{equation}
    \Sigma_{\text{new}} - \Sigma_{\text{current}} = \mathbf{k}(x)s(x)^{-1}\mathbf{k}(x)^T
\end{equation}

The term $\mathbf{k}(x)s(x)^{-1}\mathbf{k}(x)^T$ is an outer product of vectors, resulting in a rank-1 matrix. Thus, each measurement can only reduce uncertainty in a single direction in parameter space:

\begin{equation}
    \text{rank}(\Sigma_{\text{i}} - \Sigma_{\text{i+1}}) \leq 1
\end{equation}

This rank limitation highlights the importance of carefully choosing measurement actions to maximize information gain in the most relevant directions of parameter space. It also motivates the development of natural basis representations for the parameter space that align with the measurement operators, effectively reducing the problem to a finite-dimensional inference task.

\subsection{Information optimal sampling}

The Bayesian inference framework provides a principled approach for selecting measurements to maximize information gain about the unknown parameter $\theta$. Information-theoretic measures, particularly mutual information, offer a natural quantification of the value of potential measurements.

The mutual information between random variables $X$ and $Y$ is defined as:
\begin{equation}
    I(X;Y) = \mathbb{E}_{p(x,y)}\left[\log\frac{p(x,y)}{p(x)p(y)}\right] = H(X) - H(X|Y)
\end{equation}
where $H(X)$ is the entropy of $X$ and $H(X|Y)$ is the conditional entropy of $X$ given $Y$.

In our measurement context, the mutual information between the parameter $\theta$ and a potential measurement $y$ at action $x$ represents the expected reduction in uncertainty about $\theta$ after observing $y$:
\begin{equation}
    I(\theta;y|x) = H(\theta) - H(\theta|y,x)
\end{equation}

For Gaussian distributions, the (differential) entropy has a closed-form expression related to the determinant of the covariance matrix:
\begin{equation}
    H(\theta) = \frac{1}{2}\log((2\pi e)^d|\Sigma_\theta|)
\end{equation}
where $d$ is the dimensionality of $\theta$.

Consequently, the mutual information between $\theta$ and a measurement at action $x$ can be expressed as:
\begin{equation}
    I(\theta;y|x) = \frac{1}{2}\log\frac{|\Sigma_{\text{current}}|}{|\Sigma_{\text{new}}|}
\end{equation}

Using the update equations from the previous section and matrix determinant properties, we can derive:
\begin{equation}
    I(\theta;y|x) = \frac{1}{2}\log\left|I + \frac{1}{\sigma^2_\varepsilon(x)}\mathcal{A}(x)\Sigma_{\text{current}}\mathcal{A}(x)^T\right|
\end{equation}

For scalar measurements, this simplifies to:
\begin{equation}
    I(\theta;y|x) = \frac{1}{2}\log\left(1 + \frac{1}{\sigma^2_\varepsilon(x)}\mathcal{A}(x)\Sigma_{\text{current}}\mathcal{A}(x)^T\right)
\end{equation}

These two expressions directly quantify the expected information gain from a measurement at $x$, corresponding to two optimal sampling strategies:
\begin{enumerate}
    \item Global (designing a set of measurements to maximize total information gain)
    \item Myopic (choosing each next measurement to maximize immediate information gain)
\end{enumerate}

\subsubsection{Specifications leading to classical sampling}\label{sec:classical}

To complement the more intuitive explanations given in the main text, this section provides more rigorous derivations how Nyquist sampling emerges as the information optimal strategy. Specifically, we show why the signal can be expressed by a limited number of parameters, why these parameters can be expressed in a trigonometric bases, why information optimal sensing reduces to optimizing the determinant of the Fisher information matrix and how these insights yield to the condition of equidistant sampling within the optimal measurement strategy.

\paragraph{Finite Dimensionality of the Parameter Space}
\label{sec:finite_dimensionality}

When we speak of an ``unknown signal'' \(f(t)\), we may picture a function that could in principle live in an \emph{infinite}-dimensional space. As we will show below, two physically unavoidable constraints in the conjugate domains of time and frequency, however, collapse the effective dimension to a \emph{finite} number:

\begin{enumerate}
  \item \textbf{Band-limitation.} Any real electromagnetic or acoustic field is produced by sources with finite response time and therefore has a maximum relevant angular frequency \(B\) (rad$\cdot s^{-1}$ ).  Formally,
    \[
      \widehat f(\omega)=0
      \quad\text{for}\quad
      |\omega|>B.
    \]
  \item \textbf{Finite observation window.} Measurements take place during a finite time interval of length \(T\).  Outside \(\lvert t\rvert\le T/2\) the field is simply \emph{not recorded}. 
\end{enumerate}

Now those two limitations would result in a finite area in phase space which could be occupied by a finite number of \textit{logons}, the minimal resolvable phase space area resulting from the time-frequency uncertainty principle as stated by Gabor \cite{gabor1946theory}. What this simplified view masks is that a sharp mask by band-limitation results in an infinitely slow decay on the conjugate domain, so the signal cannot be time-limited. Nevertheless, Slepian, Pollak, and Landau have formalized this intuition of dimensionality using Prolate Spheroidal Wave Functions (PSWFs), which form an eigenspace of the time and frequency limitation operators and critically, whose first $\lfloor 2BT\rfloor$ eigenvalues are unity and go exponentially to zero thereafter \cite{slepian_bandwidth_1976}. This eigenanalysis was generalized to arbitrary localization operators in phase space by Daubechies, showing that e.g.\ Gaussian time and frequency limiting results in a countable set of eigenfunctions, albeit with different asymptotic behaviour of the eigenvalues \cite{daubechies1988time}.  

The key pracitcal justification for a using a discrete basis was given by Landau and Pollak by showing that any simultaneusly band-limited and approximately time-limited $f(t)$ for which 
\begin{equation}
    \frac{\int_{| t | \leq T / 2} \mid f ( t ) \mid^{2} d t} {\int_{-\infty}^{\infty} \mid f ( t ) \mid^{2} d t}=1-\epsilon_{T}^{2} 
\end{equation}
can be approximated using $2BT$ independent signals with an error proportional to the unconcentrated energy $\epsilon_T^2$ using the PSWF basis and with an error proportional to $\epsilon_T + \epsilon_T^2$ using a trigonometric basis\cite{landau_prolate_1962}. While PSWFs are the theoretically optimal choice, their computational complexity relative to the minimal error advantage makes the trigonometric basis the practical choice for real-world signal reconstruction problems.

\paragraph{Modeling the Constrained Signal: From Gaussian Processes to a Finite Trigonometric Basis}

To model an unknown signal $s(t)$ from a general non-parametric standpoint, we can begin by considering it as a realization of a Gaussian Process (GP) \cite{rasmussen2003gaussian}, $s(t) \sim \mathcal{GP}(\mu(t), \sigma(t,t'))$. Such a process is characterized by its mean function $\mu(t)$ and covariance kernel $\sigma(t,t')$.

Any GP can be expressed via a Karhunen-Lo{\`{e}}ve expansion \cite{wang2008karhunen}:
$s(t) = \mu(t) + \sum_{n=0}^{\infty} \sqrt{\lambda_n} z_n \psi_n(t)$,
where $z_n \sim \mathcal{N}(0,1)$ are uncorrelated random variables, and $\psi_n(t)$ and $\lambda_n$ are the eigenfunctions and eigenvalues, respectively, of its covariance kernel $\sigma(t,t')$.

In the preceding section, we established that a signal subjected to a finite observation window of duration $T$ and a finite bandwidth $B$ is optimally represented, in terms of energy concentration, by Prolate Spheroidal Wave Functions (PSWFs). If we construct the covariance kernel $\sigma(t,t')$ of our GP to perfectly embody these strict time and band limitations, then the eigenfunctions $\psi_n(t)$ in the Karhunen-Lo{\`{e}}ve expansion become these PSWFs. A critical property of this system is the behavior of the corresponding eigenvalues $\lambda_n$: approximately $2BT$ of these eigenvalues are close to unity, while beyond this number, they plummet precipitously towards zero. The sum in the KL expansion can thus be truncated after these $\approx 2BT$ terms with negligible loss of information.

While these PSWFs, the $\psi_n(t)$, form the natural eigenbasis for signals under the aforementioned strict constraints, a trigonometric basis (sines and cosines) is often preferred for its practical advantages, as exemplified by Shannon's foundational work on band-limited signals \cite{shannon_communication_1949}. It is also understood that for large time-bandwidth products, such a trigonometric basis can effectively represent signals within the space defined by the dominant PSWFs. Thus, we adopt a real-valued trigonometric polynomial of degree $m$ as our practical signal model:
$S(t) = a_0 + \sum_{k=1}^{m} (a_k \cos(2\pi k f_0 t) + b_k \sin(2\pi k f_0 t))$.
Here, $f_0 = 1/T$ is the fundamental frequency. This model has $p = 2m+1$ parameters. By choosing $p$ to match the signal's intrinsic dimensionality $K \approx 2BT$ derived from the PSWF analysis, we ensure $m \approx BT$. This implies that the highest frequency represented in our trigonometric model, $f_{\text{max\_poly}} = m \cdot f_0 \approx (BT) \cdot (1/T) = B$, aligns with the signal's bandlimit.

This connection to Gaussian Processes also offers a clarifying perspective on classical methods versus a full Bayesian treatment. Classical signal reconstruction via Shannon's interpolation formula, using sinc functions from Nyquist-rate samples, can be understood as analogous to obtaining a Maximum A Posteriori (MAP) point estimate. Such an estimate would arise if a Gaussian Process, endowed with a prior that perfectly enforces strict bandlimitation, were applied to the problem. Our multivariate Bayesian approach as presented in the main text, however, by modeling the coefficients of this finite-dimensional trigonometric basis as a multivariate Gaussian and rigorously propagating the full posterior covariance matrix, aligns with a full Bayesian Gaussian Process treatment. This inherently provides not just a point estimate of the signal, but also a comprehensive quantification of its uncertainty.

\paragraph{Uninformed Prior and D-Optimality}

A critical assumption that leads to classical sampling theory is an uninformed prior across parameter components. Mathematically, this corresponds to:
\begin{equation}
    \Sigma_0 = \alpha\mathbf{I}
\end{equation}
where $\alpha$ is a large constant representing high initial uncertainty about all parameter components equally.

Under this assumption, the information gain from a set of measurements becomes:
\begin{equation}
    I(\theta;\mathbf{y}|\mathbf{x}) = \frac{1}{2}\log\left|I + \frac{\alpha}{\sigma^2}\mathbf{A}\mathbf{A}^T\right|
\end{equation}

In the limit as $\alpha \to \infty$ (complete initial ignorance), maximizing information gain becomes equivalent to maximizing:
\begin{equation}
    \log|\mathbf{A}^T\mathbf{A}|
\end{equation}

This is precisely the classical D-optimality criterion from experimental design theory. The optimization problem reduces to finding the set of measurement actions $\{x_1, x_2, \ldots, x_n\}$ that maximizes the determinant of the Fisher information matrix $\mathbf{A}^T\mathbf{A}$.

\paragraph{Emergence of Nyquist Sampling}

The previous sections have established two crucial results. First, we showed that the (approximately) band- and time-limited signal can be modeled using a finite number of trigonometric bases functions. Second, in the previous section we found that complete ignorance reduces the Bayesian information gain criterion to the D-optimality criterion. Conveniently, combining these insights allows us to directly employ a theorem from experimental design theory (see Pukelsheim \cite{pukelsheim_optimal_2006}) to establish exactly that for a complete trigonometric basis of degree $m$ measured on $[0,T]$, equispaced points constitute a D-optimal design.

This result is profound because it means that uniform sampling at the Nyquist rate is not merely sufficient for signal reconstruction (as typically presented) but is actually optimal in an information-theoretic sense when no prior knowledge distinguishes the importance of different frequency components.

\paragraph{Global versus myopic optimization}

The above results show that Nyquist sampling emerges -- under specific constraints -- as unique optimal acquisition protocol to maximize information gain. 
However, for arbitrary non-uniform priors such closed-form solution are elusive and the sampling policy needs to be found numerically. As discussed above, both global and myopic approaches are applicable. The former could for instance employ evolutionary methods, but for computational efficiency (and to allow for non-linear proportional noise terms) it is more attractive to use a fully exploiting approach that maximizes information gain in the next measurement.

As we discuss in the main text, a central appeal of our method is that it reverts to conventional sampling in absence of prior knowledge. However, does this also apply to an algorithm that pursues myopic instead of global optimization? The result is that it does so \textit{approximately}: After an initial measurement extracting the DC component, the second measurement delay $\tau_2$ maximizes immediate information gain, yielding a transcendental equation for normalized delay $z = \tau_2(\omega_U - \omega_L)$:

\begin{equation}
g(z) = \frac{1}{2} + \frac{\sin(2z)}{4z} - \frac{\sin^2(z)}{z^2}
\end{equation}

The maximum occurs at $z \approx 1.164\pi$, deviating from the Nyquist interval by about 16\%. This reflects the myopic optimization without knowledge of future measurements. From the very next measurement onward, the optimal delays align almost perfectly with Nyquist intervals. Numerical analysis shows that subsequent delays differ from the Nyquist spacing by less than 1\%, effectively reproducing the classical sampling pattern. For a frequency-limited signal between $\omega_l$ and $\omega_u$, this corresponds precisely to the optimal sampling rate predicted by bandpass sampling theory.
This rapid convergence can be understood through the lens of basis transformation. Initially, the spectrum's uncertainty is uniformly distributed across frequency bins in the natural frequency basis. Each measurement effectively reveals a particular discrete cosine transform (DCT) mode, removing uncertainty along that pattern. After just a few measurements, the underlying structure of uncertainty no longer appears uniform in frequency space--instead, it becomes naturally expressed in the DCT basis. Once this basis alignment occurs, selecting delays that correspond to pure DCT modes (which exactly coincide with Nyquist intervals) becomes the mathematically optimal choice for extracting maximum information. This emergence of Nyquist sampling from the information-theoretic framework provides a powerful validation. The fact that we quickly converge to the known optimal solution in this special case suggests that its single-step optimization strategy is indeed effective at discovering optimal sampling patterns, even when performing computationally-lightweight, myopic optimization.

\subsection{Connection to Fourier Transform Spectroscopy}

In Fourier transform spectroscopy (FTS), the intensity of light is modulated as a function of time delay $\tau$ between two interfering beams. The measured fluence forms an interferogram that relates to the spectrum through the integral relationship:

\begin{equation}
F(\tau) = 2\int_{-\infty}^{\infty} |E(t)|^2dt + 2\Re\left[\int_{-\infty}^{\infty} E(t)E^*(t-\tau)dt\right]
\end{equation}

Expressing the electric field in terms of its Fourier transform, measurements take the form:
\begin{equation}
    F(\tau) = \int_\Omega S(\omega)[1 + \cos(\omega\tau)]d\omega 
\end{equation}
where $\tau$ represents the time delay between interferometer arms.

Under the standard FTS approach, delays are selected uniformly according to the Nyquist criterion
\begin{equation}
\Delta\omega_{max} = \frac{\pi}{\delta\tau}, \quad \delta\omega = \frac{2\pi}{\Delta\tau_{max}}
\label{resolution}
\end{equation}
which allows the retrieval of the spectrum by Fourier transformation of the complete interferogram. 

This approach implicitly assumes no prior knowledge about the spectrum. The derivation above provides the theoretical justification for this approach: uniform sampling at the Nyquist rate is information-optimal when operating with a uniform prior.
However, in most practical applications, considerable prior knowledge exists about the spectrum--ranging from physical constraints on the blackbody illumination to domain-specific knowledge about expected spectral features. The Active Fourier Spectroscopy approach presented in this paper capitalizes on this prior knowledge to enable more efficient and accurate spectral reconstruction.

\subsubsection{Linear probabilistic inference}
\label{sec:linear_fts}

The autocorrelation measurements used in FTS can be further generalized upon realizing that the fluence at any delay yields information about the entirety of the spectrum through the integral relationship, as opposed to e.g. a dispersive spectroscopic measurement. This opens up the possibility of inferring the spectrum in a sequential manner, updating an estimate with each new measurement.

One such measurement can be expressed in a linear equation by rewriting the integral equation on a grid of frequencies and delays as

\begin{equation}
    F_{\tau} = R_{\tau\omega}S_{\omega} \quad \mathrm{with} \quad R_{\tau\omega} = 2[1 + \cos(\omega\tau)]Q(\omega)
\end{equation}

where $Q(\omega)$ represents the detector's spectral efficiency.

Our estimate of the spectrum which encapsulates all acquired information at any point is most naturally expressed as a probability distribution. 
The formal method of updating our knowledge about the spectrum based on new measurements follows Bayes' theorem:
\begin{equation}
    p(S|F) = \frac{p(F|S)p(S)}{p(F)}
\end{equation}
where $p(S|F)$ is our updated (posterior) knowledge of the spectrum after observing fluence $F$, $p(F|S)$ represents how likely we are to observe fluence $F$ given spectrum $S$, and $p(S)$ encodes our prior knowledge.

In practice, each measurement includes noise, which we initially model as additive Gaussian:
\begin{equation}
    F(\tau) = I_0(\tau) + n_{add}(\tau), \quad n_{add}(\tau) \sim \mathcal{N}(0,\sigma^2_{add})
\end{equation}
This noise model leads to a Gaussian likelihood function:
\begin{equation}
    p(F|S) \propto \exp\left(-\frac{(F - RS)^2}{2\sigma^2_{add}}\right)
\end{equation}

Several arguments motivate representing our knowledge of the spectrum as a Gaussian distribution. First, many physical processes naturally lead to Gaussian distributions through the Central Limit Theorem. Second, given only constraints on mean and variance, the Gaussian maximizes entropy, making it the least presumptive choice \cite{jaynes2003probability}. Further, from a practical point of view, Gaussian distributions are self-conjugate and when both prior and likelihood are Gaussian, the posterior is also Gaussian, enabling analytical solutions.

Representing our prior knowledge as a Gaussian with mean $\mu_S$ and covariance $\Sigma_S$, its distribution is:
\begin{equation}
    p(S) \propto \exp\left(-\frac{1}{2}(S-\mu_{old})^T\Sigma_{old}^{-1}(S-\mu_{old})\right)
\end{equation}

To derive the update equations, we examine the product of likelihood and prior. Taking the exponents:
\begin{equation}
\begin{split}
    -\frac{1}{2}(S-\mu_{old})^T\Sigma_{old}^{-1}(S-\mu_{old}) - \frac{1}{2\sigma^2_{add}}(F - RS)^2
\end{split}
\end{equation}
Expanding the quadratic terms:
\begin{equation}
\begin{split}
    -\frac{1}{2}&[S^T\Sigma_{old}^{-1}S - 2\mu_{old}^T\Sigma_{old}^{-1}S + \mu_{old}^T\Sigma_{old}^{-1}\mu_{old} \\
    &+ \frac{1}{\sigma^2_{add}}(S^TR^TRS - 2FRS + F^2)]
\end{split}
\end{equation}
Collecting terms in $S$:
\begin{equation}
\begin{split}
    -\frac{1}{2}&[S^T(\Sigma_{old}^{-1} + \frac{1}{\sigma^2_{add}}R^TRS \\
    &- 2({\mu_{old}^T\Sigma_{old}^{-1}} + \frac{F}{\sigma^2_{add}}R)S + \text{const.}]
\end{split}
\end{equation}
This quadratic form in $S$ must represent a Gaussian. Comparing to the standard form, we can identify:
\begin{equation}
    \Sigma_{new}^{-1} = \Sigma_{old}^{-1} + \frac{1}{\sigma^2_{add}}R^TR
\end{equation}
Using the Woodbury matrix identity:
\begin{equation}
    \Sigma_{new} = \Sigma_{old} - \Sigma_{old}R^T(R\Sigma_{old}R^T + \sigma^2_{add})^{-1}R\Sigma_{old}
\end{equation}
This can be rewritten as:
\begin{equation}
    \Sigma_{new} = (I - \gamma R)\Sigma_{old}
\end{equation}
where $\gamma = \Sigma_{old}R^T(R\Sigma_{old}R^T + \sigma^2_{add})^{-1}$
Similarly, from the linear terms in $S$, the new mean must satisfy:
\begin{equation}
    \Sigma_{new}^{-1}\mu_{new} = \Sigma_{old}^{-1}\mu_{old} + \frac{F}{\sigma^2_{add}}R^T
\end{equation}
Substituting the expression for $\Sigma_{new}$ and simplifying yields:
\begin{equation}
    \mu_{new} = \mu_{old} + \gamma(F - R\mu_{old})
\end{equation}
These update equations maintain Gaussian form while incorporating new measurements optimally.
Incidentally, for a zero mean prior with no prior knowledge (represented by a very large diagonal prior covariance) this equation simplifies to $\mu_{new} = R^{-1}F$. Because Nyquist sampling is information optimal in this case, $R$ turns out to be the real part of a DCT matrix plus a DC term. This further explains FTS as a limiting case of AFS. 

In reality, the measurement noise contains both additive and multiplicative components:
\begin{equation}
    F(\tau) = I_0(\tau)(1 + n_{mult}(\tau)) + n_{add}(\tau)
\end{equation}
where $n_{mult}(\tau) \sim \mathcal{N}(0,\sigma^2_{mult})$. For small noise levels, the product term $n_{add}n_{mult}$ is second-order small and can be approximated as Gaussian. The total noise contribution becomes:
\begin{equation}
    n(\tau) = I_0(\tau)n_{mult}(\tau) + n_{add}(\tau) + n_{prod}(\tau)
\end{equation}
with $n_{prod}(\tau) \sim \mathcal{N}(0,\sigma^2_{add}\sigma^2_{mult})$. Since the sum of independent Gaussian terms remains Gaussian, the total noise variance is:
\begin{equation}
    \text{Var}[n(\tau)] = I_0(\tau)^2\sigma^2_{mult} + \sigma^2_{add} + I_0(\tau)^2\sigma^2_{add}\sigma^2_{mult}
\end{equation}
This maintains the analytical tractability of the update equations while capturing both noise contributions.

\subsection{Prior knowledge incorporation}

Effective incorporation of prior knowledge is crucial to Bayesian inference, distinguishing our approach from traditional sampling techniques. In this section, we describe how physically motivated, sample-specific constraints and general structural assumptions can be formally translated into rigorous prior distributions. Equally important is the principled selection and weighting of these priors through Bayesian model evidence, ensuring that the inference process remains both robust and unbiased by potential prior misspecification.

\subsubsection{Prior construction}
Prior knowledge about spectra to be measured generally falls into two distinct categories. The first category comprises sample-specific priors that encode physical constraints and known spectral characteristics. For example, in FTIR spectroscopy of blood plasma, the spectrum is constrained both by the blackbody illumination profile and typical biological absorption features. These priors effectively reduce the dimensionality of the inference problem by restricting the search space to physically plausible solutions.

The second category consists of general structural priors that encode assumptions about spectral properties independent of the specific sample. A fundamental example is spectral smoothness - the assumption that nearby frequency components exhibit correlations. Such structural priors are essential for making the Bayesian inverse problem well-posed\cite{latz_bayesian_2023}, as they provide regularization through the introduction of appropriate prior measures.

A canonical example of structural prior construction is the radial basis function (RBF) covariance, which encodes smoothness assumptions through frequency correlations. The RBF covariance between frequencies $\omega_i$ and $\omega_j$ takes the form:
\begin{equation}
    \Sigma_{ij} = \sigma^2\exp\left(-\frac{(\omega_i-\omega_j)^2}{2l^2}\right)
\end{equation}
where $l$ is the correlation length and $\sigma^2$ the overall variance scale. This form arises from assuming the spectrum is a Gaussian process with stationary covariance that depends only on frequency differences. The correlation length $l$ determines the characteristic scale over which spectral features vary, effectively imposing a smoothness constraint. The construction follows from requiring (a) \textit{Stationarity} ($\text{Cov}(\omega_i,\omega_j) = k(|\omega_i-\omega_j|)$ for some function $k$), (b) \textit{Smoothness} ($k$ is differentiable), (c) \textit{Normalization} ($k(0)=\sigma^2$) and (d) \textit{Positive definiteness} (all eigenvalues of $\Sigma$ are positive). The RBF form is the simplest covariance satisfying these conditions, though other choices like Mat{\'{e}}rn kernels are possible for different smoothness assumptions.

It should be noted that spectral correlation is not only a property of the sample, but also of the measurement device. Spectral resolution is classically determined by the maximum delay, $\delta\omega = \frac{2\pi}{\tau_{\text{max}}}$. This resolution limit can be interpreted as two frequencies becoming distinguishable when they accumulate sufficient phase difference over the measurement range. In AFS with an RBF kernel, we can encode equivalent resolution constraints through the correlation length parameter $l$. Specifically, when comparing the instrument line shape function of classical FTS (a sinc function) with the proposed Gaussian correlation structure of the RBF kernel, we need to determine when two spectral components should be considered "resolved" in each framework. In classical FTS, frequencies separated by $\delta\omega$ produce a full cycle of phase difference over $\tau_{\text{max}}$. In the RBF framework, frequencies are effectively distinguishable when their correlation drops below a certain threshold. By matching these criteria and considering the width of the central lobe of the sinc function, we find that $l \approx \frac{\pi}{\tau_{\text{max}}} = \frac{\delta\omega}{2}$ provides an equivalent resolving power. This relationship allows us to translate classical resolution constraints into probabilistic prior constraints in a principled manner.

Beyond these general structural priors, we can incorporate specific spectral features of interest by locally modifying the covariance structure. For instance, when absorption lines are expected at certain frequencies, we can increase the prior variance in these regions:
\begin{equation}
    \Sigma_{ii} = \sigma^2(1 + \alpha\exp(-(\omega_i-\omega_0)^2/2w^2))
\end{equation}
where $\omega_0$ is the expected line position, $w$ its width, and $\alpha$ controls the increased uncertainty. This construction maintains the positive definiteness of the covariance while allowing greater flexibility in regions of interest.

The combination of sample-specific and structural priors serves complementary purposes. Sample-specific priors provide efficient inference by constraining solutions to physically relevant subspaces, while structural priors ensure mathematical well-posedness through appropriate regularization. Both are essential for robust spectral reconstruction. The well-posedness follows from the fact that the posterior inherits the regularity of the prior measure\cite{latz_bayesian_2023}, making the inverse problem stable with respect to perturbations in the data.

\subsubsection{Prior selection}\label{section:prior_selection}
The transition from traditional FTS to AFS enables significant enhancement through the incorporation of prior information. However, a rigorous framework is needed to ensure that prior information does not lead to biases that deteriorate performance. The challenge lies in selecting or averaging over a set of candidate priors (henceforth called prior models) in a way that optimizes the spectral reconstruction while avoiding overfitting.

The orthodox Bayesian approach would consider a continuous distribution over possible prior models and marginalize them:

\begin{equation}
p(S|F) = \int p(S|F,M)p(M|F)dM
\label{eq:prior_marginalization}
\end{equation}

where $S$ is the spectrum, $F$ is the data, and $M$ represents a prior model. However, this integral is often intractable to calculate analytically and computationally expensive to approximate using methods such as Markov Chain Monte Carlo (MCMC). To make the problem tractable, we consider a discrete set of candidate prior models $\{M_1, M_2, ..., M_K\}$. 

Assuming equally likely prior models, calculating $p(M_k|F)$ amounts to calculating the evidence $p(F|M_k)$ . In our setting, where the fluence data are accumulated in increments \(F_1, F_2, \dots, F_n\), the evidence can be expressed via sequential factorization:
\begin{equation}
p(F \mid M_k) 
\;=\;
\prod_{i=1}^n 
p\bigl(F_i \,\big|\,
       F_1,\dots,F_{i-1},\,M_k\bigr).
\label{eq:seq_evidence_factorization}
\end{equation}
Each factor 
\(p(F_i \mid F_{1..i-1},M_k)\)
can be interpreted as the one-step-ahead predictive probability for the \(i\)-th measurement -- i.e.\ how well the model predicts new data that were not yet included in the posterior. In that sense, Eq.~\eqref{eq:seq_evidence_factorization} conceptually ``leaves out'' future measurements from the posterior used for predicting the next measurement, which is reminiscent of ``leave-one-out'' cross-validation ideas. This illustrates how marginalization evaluates not only the fit of the model but also predictive (generalization) capabilities. The cumulative sum of the marginal log-likelihood of new data under the current prior is thus directly the log-evidence term. This fact allows for computationally cheap updating steps in any numerical implementation.

The mentioned trade-off between fit and generalization can be directly seen in the form of the evidence
\begin{equation}
    \log p(F|M_k) = -\frac{1}{2} \bigl( (F-R\mu_S)^{\intercal}(\Sigma_F + R\Sigma_SR^{\intercal})^{-1}(F-R\mu_S) 
    + \log\det{(\Sigma_F + R\Sigma_SR^{\intercal})}\bigr)
\end{equation}
where $\mu_S, \Sigma_S$ are the moments of the prior and $\Sigma_F$ is the noise covariance. The first term can be understood as a penalization on the deviation of the data from its expectation under the prior whereas the determinant term would constitute a penalization on the model complexity. 

Having obtained the proper normalization (via the evidence term) for each model, the final posterior is a combination of each of these candidate posteriors. Again, the combined distribution has to be normalized, which can be done by weighting models using a softmax function with temperature parameter $T$:

\begin{equation}
w_k = \frac{\exp(-L_k/T)}{\sum_{j=1}^K \exp(-L_j/T)}
\label{eq:softmax_weights}
\end{equation}

where $L_k$ is a shorthand for the cumulative sum of the sequentially computed marginal log-likelihoods. As $T \to 0$, this approaches the hard selection of the best model, while higher T values produce a more uniform weighting. While this is not strictly included in \ref{eq:prior_marginalization}, it can be motivated by the fact that we are summing over a discrete set of models with potentially nonuniform coverage of the possible prior space, which makes selection more robust than averaging.

\subsubsection{Design selection under mixture priors}
\label{sec:mixture_ig}

The preceding sections established how to construct multiple candidate priors and how to weight them based on accumulated evidence. Here we address a complementary question: how should the information gain for design selection be computed when operating with a mixture of Gaussian priors?

\paragraph*{Exact formulation.}
The information gain from a measurement at delay $\tau$ is defined as the mutual information between the spectrum $S$ and the observation $F$:
\begin{equation}
    \mathrm{IG}(\tau) = I(S;F|\tau) = H(S) - H(S|F,\tau).
\end{equation}
This can equivalently be written as an expectation over the joint distribution:
\begin{equation}
    \mathrm{IG}(\tau) = \int p(S,F|\tau) \log \frac{p(S|F,\tau)}{p(S)}\,dS\,dF.
\end{equation}
For a single Gaussian prior, both the prior and posterior entropies have closed forms involving covariance determinants, yielding the analytical expression in Eq.~\eqref{eq:information_gain} of the main text. For a mixture prior $p(S) = \sum_k w_k \mathcal{N}(S;\mu_k,\Sigma_k)$, however, neither the prior entropy nor the posterior entropy admits a closed-form expression, and the information gain must in principle be estimated via Monte Carlo sampling.

\paragraph*{Decomposition into weighted and model-selection terms.}
To understand when closed-form approximations suffice, we introduce the model index $k$ explicitly into the generative model:
\begin{equation}
    p(k,S,F|\tau) = p(k)\,p(S|k)\,p(F|S,\tau),
\end{equation}
where $p(k) = w_k$ and $p(S|k) = \mathcal{N}(S;\mu_k,\Sigma_k)$. Using the chain rule for mutual information and noting that $I(k;F|S,\tau) = 0$ (the measurement likelihood depends only on $S$, not on which model generated it), we obtain:
\begin{equation}
    I(S;F|\tau) = I(S;F|k,\tau) + I(k;F|\tau).
    \label{eq:ig_decomposition}
\end{equation}
The first term is the \emph{weighted-average information gain}:
\begin{equation}
    I_{\mathrm{weighted}}(\tau) \equiv I(S;F|k,\tau) = \sum_k w_k \, I(S;F|k{=}k,\tau),
\end{equation}
which is precisely a weighted sum of the closed-form information gains for each individual model component. The second term captures the \emph{model-discriminating information}:
\begin{equation}
    I_{\mathrm{model}}(\tau) \equiv I(k;F|\tau) = H(w) - \mathbb{E}_{F|\tau}\bigl[H(w'(F))\bigr],
\end{equation}
where $w'(F)$ denotes the updated model weights after observing $F$, and $H(w) = -\sum_k w_k \log w_k$ is the entropy of the weight distribution. Since conditioning cannot increase entropy, the updated weights satisfy $H(w'(F)) \leq H(w)$, yielding the bound:
\begin{equation}
    0 \leq I_{\mathrm{model}}(\tau) \leq H(w).
    \label{eq:imodel_bound}
\end{equation}

\paragraph*{When the weighted approximation suffices.}
The decomposition in Eq.~\eqref{eq:ig_decomposition} shows that our weighted-average approximation neglects $I_{\mathrm{model}}(\tau)$. This term affects the \emph{ranking} of candidate delays only if some measurements discriminate between models more effectively than others -- that is, if the predictive distributions $p(F|\tau,k) = \mathcal{N}(F; R(\tau)\mu_k, R(\tau)\Sigma_k R(\tau)^\top + \sigma_n^2)$ differ substantially across models in a $\tau$-dependent manner. Three limiting cases justify the approximation:

\begin{enumerate}
    \item \textbf{Identical components.} If all mixture components share the same moments, the mixture reduces to a single Gaussian and the closed-form expression is exact.

    \item \textbf{Similar means, different covariances.} When the prior means $\mu_k$ are approximately equal but covariances $\Sigma_k$ differ, the predictive means $R(\tau)\mu_k$ coincide for all $\tau$. The mixture of predictive distributions can then be well-approximated by a single Gaussian with covariance computed from the mixture moments, and the model-selection term contributes negligibly to the ranking of delays.

    \item \textbf{Concentrated weights.} When the evidence strongly favors one model (i.e., $w_k \approx 1$ for some $k$), we have $H(w) \to 0$, and the bound in Eq.~\eqref{eq:imodel_bound} ensures $I_{\mathrm{model}}(\tau) \approx 0$ regardless of how different the predictive distributions may be.
\end{enumerate}

In practice, our demonstrations fall into case (3): after the first few measurements, accumulated evidence typically concentrates the weights on one or two well-performing models, rendering the approximation effectively exact. Furthermore, the temperature parameter $T$ in the softmax weighting (Eq.~\eqref{eq:softmax_weights}) provides explicit control: setting $T \to 0$ implements hard model selection, which recovers the exact information gain for the dominant model.

\paragraph{Computational considerations.}
When none of the above conditions hold and the full mixture entropy must be estimated, Monte Carlo sampling is required. The numerical stability of such estimates depends on the formulation. Direct sampling from $p(S)$ and evaluating the log-probability ratio $\log[p(S|F,\tau)/p(S)]$ can encounter numerical issues when mixture components have disparate scales or limited overlap, as samples from one component may have negligible probability under others. An equivalent but more stable formulation uses:
\begin{equation}
    \mathrm{IG}(\tau) = H(F|\tau) - H(F|S,\tau),
\end{equation}
where $H(F|S,\tau) = \frac{1}{2}\log(2\pi e \sigma_n^2)$ depends only on the measurement noise and is independent of the prior. The predictive entropy $H(F|\tau)$ involves the mixture:
\begin{equation}
    p(F|\tau) = \sum_k w_k \mathcal{N}\bigl(F; R(\tau)\mu_k,\, R(\tau)\Sigma_k R(\tau)^\top + \sigma_n^2\bigr),
\end{equation}
which is numerically better-behaved as it operates in the scalar observation space rather than the high-dimensional parameter space. It is however also possible to calculate the closed form \textit{weighted-average} term and sample the \textit{model-discriminating} term, which has the advantage of lower sampling standard error. 
In general, this calculation still requires orders of magnitude longer computation time. This computational gap implies a measurement-rate-dependent trade-off: for instruments with fast acquisition, the time spent computing the fully optimal design could instead be spent acquiring additional measurements. In such regimes, the ``good enough'' approximation combined with more measurements may outperform the exact optimum with fewer measurements.

\subsection{Uncertainty estimation}

\subsubsection{Properties of uncertainty estimates and Fellgett's advantage}
The posterior covariance matrix $\Sigma_S$ provides quantitative uncertainty estimates for reconstructed spectra. For a single spectral component under constant measurement uncertainty:

\begin{equation}
\sigma^2_{posterior} = (1-\gamma)\sigma^2_{prior}
\end{equation}

where $\gamma = \sigma^2_{prior}/(\sigma^2_{prior} + \sigma^2_{meas})$ is the update gain. As $\sigma^2_{prior} \to \infty$, representing complete initial ignorance, we recover $\sigma^2_{posterior} \approx \sigma^2_{meas}$. For n identical measurements, uncertainty reduces as $\sigma^2_{posterior} \approx \sigma^2_{meas}/n$, consistent with the central limit theorem.

For M measurements with delays uniformly distributed over ranges much larger than characteristic oscillation periods, the average measurement matrix approaches unity. This demonstrates Fellgett's advantage: variance in each spectral component scales as 1/M, providing multiplex advantage over scanning monochromator spectrometers. This advantage holds when measurement noise is not photon-dominated.

\subsubsection{Noise considerations for FTIR simulations}
Realistic simulation of FTIR measurements requires accurate modeling of time-domain noise characteristics. Commercial FTIR instruments typically provide only processed absorbance spectra rather than raw interferograms, limiting direct access to time-domain noise statistics and none of a number of contacted manufacturers could provide us with an authoritative estimate. To nevertheless ensure a meaningful comparison of FTS and AFS methods, we used Quality Control (QC) data from the L4L study -- designed to contain no biological variation -- as an empirical basis. We implemented a reverse-engineering approach in the following way:
\begin{itemize}
    \item Converting QC absorbances to spectral intensity using a blackbody illumination model (see Methods)
    \item Computing the sample covariance and transforming it to the time domain through the measurement operator
    \item Extracting the mean of diagonal elements as white noise for simulations. This amounted to $\sim\SI{99}{\dB}$ time-domain SNR, which we deemed as realistic given reported spectral SNR reported in many instrument brochures.
\end{itemize}

The apparent performance difference between FTS and AFS partly reflects standard FTIR preprocessing procedures like apodization or Savitzky-Golay smoothing (see e.g.\ \cite{baker_using_2014, morais_tutorial_2020}). These techniques effectively implement implicit spectral priors (e.g.\ smootness constraints) but in a fixed, non-adaptive manner, whereas AFS inherently formalizes such constraints explicitly with the key advantages of transparant quantification in the prior model and adaptive refinement based on evidence 

\subsubsection{Propagation of uncertainty estimates for the Vortex measurements}
For phase reconstruction, uncertainties propagate through multiple steps. AFS reconstruction yields a hyperspectral datacube $S(x,y,\lambda)$ with associated uncertainties. While spatial pixels are processed independently, the spectral channels maintain correlations through the AFS covariance matrices.
For phase gradient retrieval, we calculate the centroid positions $r(x_i,y_i,\lambda)$ for each sub-aperture at location $(x_i,y_i)$. The phase gradient is then
\begin{equation}
\nabla\phi(x_i,y_i,\lambda) = [r^{\mathrm{vortex}}(x_i,y_i,\lambda)-r^{\mathrm{ref}}(x_i,y_i,\lambda)] \frac{k(\lambda)}{f_{MLA}}
\end{equation}
Assuming only first-order effects of the intensity uncertainties on the gradient calculation we can propagate the covariance using the Delta method (see e.g. Casella \& Berger \cite{CasellaBerger2002}) as 
\begin{equation}
Cov[\nabla\phi] = J_{\nabla\phi}(S)Cov[S]J_{\nabla\phi}(S)^{\intercal}    
\end{equation}
where $J$ is the Jacobian of the gradient calculation. This approximation is motivated by its low computational complexity. We have corroborated it by comparing the resulting covariance of the phase gradients with a numerical calculation by Monte-Carlo sampling.

For phase reconstruction we first fit ideal vortex derivatives to the phase gradient maps in each wavelength channel, yielding the wavelength dependent vortex charge $l(\lambda)$, and then perform a zonal reconstruction of the residual. As the latter is a linear operation, the covariance on the phase map can be calculated in closed form. This provides comprehensive uncertainty quantification throughout the analysis chain.

\section{Extension to other applications}
\label{sec:other_apps}

The information-optimal framework applies verbatim for any linear measurement that acts in a conjugate basis. In that case, we can write
\begin{equation}
    \mathbf{y}=\mathbf{A}\mathbf{x}+\boldsymbol{\varepsilon},\qquad
    \boldsymbol{\varepsilon}\sim\mathcal{N}(\mathbf{0},\Sigma_\varepsilon),\qquad
    \mathbf{x}\sim\mathcal{N}(\mu_0,\Sigma_0),
    \label{eq:otherapps_linear}
\end{equation}
and use the same sequential Gaussian updates as introduced earlier. With a flat prior and homoskedastic noise, symmetry implies that uniform sampling in the conjugate domain is D-optimal; informative priors or heteroskedastic noise break this symmetry and the optimal design adapts. Below we explain in more detail how the formalism developed for AFS generalizes to the examples of Optical Coherence Tomography and X-ray Computed Tomography.

\subsection{Optical Coherence Tomography (OCT)}
\label{sec:OCT}

OCT is a one-dimensional analogue of AFS in the pair \(k\leftrightarrow z\). With wavenumber \(k=2\pi/\lambda\) and depth reflectivity \(r(z)\),
\begin{equation}
E_s(k)=\int r(z)\,e^{-i2kz}\,dz,\qquad
I(k)\approx |E_0(k)|^2+2\,\mathrm{Re}\!\left\{E_0(k)\int r(z)e^{-i2kz}\,dz\right\},
\label{eq:oct_field_linear}
\end{equation}
where \(E_0(k)\) is the reference spectrum. After removing the DC term, we can discretize on $z_i$ and $k_j$. Identifying $\mathbf{x}$ with $r_i$ and the matrix elements
\begin{equation}
A_{ji}=2\,\mathrm{Re}\!\bigl\{E_0(k_j)\,e^{-i2k_j z_i}\bigr\}\,\Delta z .
\label{eq:oct_A}
\end{equation}
we then recover a linear model that is algebraically identical to the one of AFS. The posterior mean and covariance update as before, and the expected information gain of a new spectral sample follows from the same rank-1 (or rank-\(d\)) update. In the ignorance limit with flat \(E_0(k)\) and homoskedastic noise, maximizing \(\log\det(\mathbf{A}^{\!\top}\mathbf{A})\) recovers uniform spacing in \(k\), i.e. the classical uniform-\(k\) requirement.

Most contemporary systems measure spectra with a camera spectrometer rather than scanning delay. The Fourier formulation remains directly useful because axial resolution is determined by the effective \(k\) grid. In Bayesian OCT this grid is encoded in \(\mathbf{A}\) via the known spectral responses of the detector; the posterior covariance reveals which radial frequencies along \(z\) are under-resolved. A small set of additional Fourier-type measurements -- e.g. few controlled delay steps -- appends informative rows to \(\mathbf{A}\) at precisely those missing \(k\) locations. This is the same algebra used in the hybrid hyperspectral imaging demonstration: coarse filter channels are represented in \(\mathbf{A}\), and a minimal interferometric supplement completes the high-frequency spectral content. Within the same inference, OCT thus uses Fourier measurements as a targeted resolution upsampler for spectrometer data, while providing full uncertainty on \(r(z)\).

\subsection{X-ray computed tomography (CT)}
\label{sec:CT}

CT extends the same logic to two-dimensional slices. Line integrals of the attenuation field \(x(\mathbf{r})\) at view angle \(\alpha\) and detector coordinate \(s\) give \(\mathbf{y}=\mathbf{A}\mathbf{x}+\boldsymbol{\varepsilon}\), with rows of \(\mathbf{A}\) implementing rays. Filtered backprojection (FBP),
\begin{equation}
\hat{\mathbf{x}}_{\rm FBP}=\mathcal{F}^{-1}\!\left\{\mathcal{F}\!\left[\mathbf{A}^{\!\top}\mathbf{y}\right]\,W(k)\right\},
\label{eq:fbp}
\end{equation}
corresponds to the ignorance limit -- an uninformed prior, equiangular sampling, and a fixed frequency weighting \(W\) -- in a way analogous to how the one-dimensional Fourier transform in the FTS case.

To make the conjugate structure explicit, expand the slice in cylindrical harmonics,
\(x(r,\theta)=\sum_{m} f_m(r)\,e^{i m\theta}\). Take the 1D Fourier transform
of each projection in the detector coordinate \(s\), and denote this transform simply by
\(P(k,\alpha)\). The Fourier--slice (central-slice) theorem in polar coordinates then gives
\begin{equation}
P(k,\alpha)=\sum_{m} i^{-m} e^{i m \alpha}\,H_m\{f_m\}(k),\qquad
H_m\{f\}(k)=\int_{0}^{\infty} f(r)\,J_m(k r)\,r\,dr,
\label{eq:polar_factorization}
\end{equation}
where \(k\ge 0\) is the radial spatial frequency and \(J_m\) is the Bessel function of the first kind.
For a fixed \(k\), the forward model can then be written as \(\mathbf{P}(k)=V\,\mathbf{c}(k)\), where the entries at view angles \(\alpha_j\) read
\[
P(k,\alpha_j)=\sum_m V_{jm}\,c_m(k),\qquad \text{with } V_{jm}=e^{i m \alpha_j},\quad
c_m(k)=i^{-m}H_m\{f_m\}(k).
\]
When all views have the same reliability (uniform spacing and noise), the least-squares solution reduces to the Moore-Penrose pseudoinverse, and the normal operator is \(N(k)=V^{*}V\), where $V^*$ is the conjugate transpose. For equiangular views one has \(V^{*}V\propto I\), so \(N(k)\) is diagonal. In the continuum limit this is precisely the setting assumed by filtered backprojection (FBP): uniform angular coverage renders the backprojection--projection operator rotationally invariant, so it is diagonal both in the cylindrical-harmonic (Hankel) representation and in the Cartesian two-dimensional Fourier representation. In that ideal regime the two descriptions are equivalent.

If angular sampling or per-angle reliability departs from uniformity, it is natural to include a (typically diagonal) weight matrix \(W(k)\) over the view angles. The \textit{weighted} normal operator $N(k)=V^{*}W(k)V$ then acquires off-diagonal terms that couple different angular orders \(m\). In Cartesian Fourier coordinates this appears as anisotropic coverage of the 2D frequency plane: the normal operator loses rotational symmetry and can no longer be diagonalized, producing familiar reconstruction artifacts like streaks and directional blur. The harmonic Hankel representation, however, maintains explicit visibility of this anisotropy through \(N(k)\), directly revealing which orders \(m\) are weakly constrained and how additional views at specific angles \(\alpha\) would modify \(N(k)\) to reduce uncertainty.

This basis choice matters for information-optimal design, especially in sparse-view regimes. Across different \(k\) the problem decouples, and -- when views are equiangular -- the angular orders \(m\) decouple as well. As a result, information gain and uncertainty reduction can be evaluated mode-by-mode in physically meaningful coordinates. Representing \(x\) on a Cartesian pixel grid, by contrast, mixes these modes: the same operator becomes dense and poorly conditioned, and the normal operator no longer
exposes which directions are unconstrained. If expected information is then computed  with diagonal posterior approximations, as in the numerical experiments of Burger et al.\ \cite{burger2021sequentially}, the optimization can be steered by pixel-grid artifacts rather than by genuine object degrees of freedom. This problem does not occur if one were to retain the full covariance matrix, but this is often computationally prohibitive. For sparse, adaptive design it is thus essential for evaluating information gain and reliable measurement optimization to work in conjugate coordinates.

\end{document}